\documentclass[aps,groupedaddress,showpacs,floatfix]{revtex4}
\usepackage{amssymb}
\usepackage{amsmath}
\usepackage{graphicx}
\DeclareMathOperator{\sgn}{sgn}
\DeclareMathOperator{\Ai}{Ai}
\begin{document}

\title{Attempted Bethe anzats solution for one-dimensional 
directed polymers in random media}

\author{Victor Dotsenko$^{\, a,c}$, Boris Klumov$^{\, b}$  $\; \; (*)$  }

\affiliation{$^a$LPTMC, Universit\'e Paris VI, 75252 Paris, France}

\affiliation{$^b$Max-Planck-Institute fur Extraterrestrische Physik, 
D-85741 Garching, Germany}

\affiliation{$^c$L.D.\ Landau Institute for Theoretical Physics,
   119334 Moscow, Russia}

\date{\today}

\begin{abstract}

We study the statistical properties of one-dimensional directed polymers in a
short-range random potential by mapping the
replicated problem to a many body quantum boson system with attractive
interactions.  We find the full set of eigenvalues and eigenfunctions of the
many-body system and perform the summation over the entire spectrum of excited
states. The analytic continuation of the obtained exact expression
for the replica partition function from integer to non-integer replica parameter $N$ 
turns out to be ambiguous. Performing the analytic continuation simply by
assuming that the parameter $N$ can take arbitrary complex values, 
and going to the thermodynamic limit of the original directed polymer
problem, we obtain the explicit universal expression for the probability
distribution function of free energy fluctuations.

\end{abstract}


\maketitle


\medskip

\begin{center}
 
{\bf \large Contents}

\end{center}

\begin{tabbing}
 AAAAAA \= {\bf Appendix C: Wave functions of quantum bosons with attractive interactions} \=
 . . . . \=  13\kill

\> {\bf I. Introduction}                             \>                   . . . .    \> 1  \\
\> {\bf II. Mapping to quantum bosons}                \>                  . . . .    \> 4  \\
\> {\bf III. Eigenstates of the one-dimensional quantum bosons system} \> . . . .    \> 5  \\
\> {\bf IV. Replica partition function}                               \> . . . .    \> 9  \\
\> {\bf V. Thermodynamic limit}                                        \> . . . .    \> 12 \\
\> {\bf VI. Free energy distribution function}                         \> . . . .    \> 13 \\
\> {\bf VII. Discussion}                                               \> . . . .    \> 15 \\
\\
\> {\bf Appendix A: Wave functions of quantum bosons with repulsive interactions}  \> . . . . \> 17  \\
\> {\bf Appendix B: Ground state of quantum bosons with attractive interactions}   \> . . . . \> 19  \\
\> {\bf Appendix C: Wave functions of quantum bosons with attractive interactions} \> . . . .  \> 21 \\
\> {\bf Appendix D}       \> . . . . \> 27  \\
\> {\bf Appendix E}       \> . . . . \> 29  \\
\> {\bf Appendix F}       \> . . . . \> 30  \\
\end{tabbing}

\section{Introduction}

Directed polymers in a quenched random potential have 
been the subject of intense investigations during the past two
decades \cite{hh_zhang_95}.
Diverse physical systems such as domain walls in magnetic films
\cite{lemerle_98}, vortices in superconductors \cite{blatter_94}, wetting
fronts on planar systems \cite{wilkinson_83}, or Burgers turbulence
\cite{burgers_74} can be mapped to this model, which exhibits numerous
non-trivial features deriving from the interplay between elasticity and
disorder.
The best understanding has been achieved for a string confined to a plane.
In this case we deal with an elastic string  directed along the $\tau$-axis 
within an interval $[0,L]$. Randomness enters the problem 
through a disorder potential $V[\phi(\tau),\tau]$, which competes against 
the elastic energy.  The problem is defined by the Hamiltonian
\begin{equation}
   \label{bas1-1}
   H[\phi(\tau), V] = \int_{0}^{L} d\tau
   \Bigl\{\frac{1}{2} \bigl[\partial_\tau \phi(\tau)\bigr]^2 
   + V[\phi(\tau),\tau]\Bigr\};
\end{equation}
where in the simplest case the disorder potential $V[\phi,\tau]$ 
is Gaussian distributed with a zero mean $\overline{V(\phi,\tau)}=0$ 
and the $\delta$-correlations: 
\begin{equation}
   \label{bas1-2}
{\overline{V(\phi,\tau)V(\phi',\tau')}} = u \delta(\tau-\tau') \delta(\phi-\phi')
\end{equation}
Here the parameter $u$ describes the strength of the disorder.
Historically, the problem of central interest was the scaling behavior of the 
polymer mean squared displacement  which in the thermodynamic limit
($L \to \infty$) is believed to have a universal scaling form
\begin{equation}
   \label{bas1-3}
\overline{\langle\phi^{2}\rangle}(L) \propto L^{2\zeta}
\end{equation}
(where $\langle \dots \rangle$ and $\overline{(\dots)}$ denote 
the thermal and the disorder averages), with $\zeta$ the so-called wandering exponent. 
More general problem for all directed polymer systems of the type, Eq.(\ref{bas1-1}),
is the statistical properties  of their free energy {\it fluctuations}. Besides the usual extensive (linear in $L$) self-averaging part $f_{0} L$ 
(where $f_{0}$ is the linear free energy density),
the total free energy $F$ of such systems contains disorder dependent 
fluctuating contribution $\tilde{F}$, which is characterized by
non-trivial scaling in $L$. It is generally believed (at list in the systems with short-ranged correlations) that in the limit of large $L$ 
the typical value of the free energy fluctuations scales with $L$ as
\begin{equation}
   \label{bas1-4}
\tilde{F}  \propto L^{\omega},  
\end{equation}
i.e. they are characterized by a single universal exponent $\omega < 1$.
In other words, in the limit of large $L$ the total (random) free energy of the system 
can be represented as
\begin{equation}
\label{bas1-5}
F \; = \; f_{0} L \; + \; \tilde{f} \; L^{\omega}
\end{equation}
where $\tilde{f}$ is the random quantity which in the thermodynamic limit $L\to\infty$ 
is described by a non-trivial universal
distribution function $P_{*}(\tilde{f})$. The derivation of this function for the system with $\delta$-correlated
random potential, Eqs.(\ref{bas1-1})-(\ref{bas1-2}) is the central issue of the present work.

One can easily note that the above two exponents $\zeta$ and $\omega$
are not independent. Indeed, since the free energy fluctuation $\tilde{F} \sim L^{\omega}$ 
can be estimated by the typical value of the elastic energy, 
$\tilde{F}  \propto \phi^{2}/L$, where, according to Eq.(\ref{bas1-3}),
the typical deviation $\phi \sim L^{\zeta}$, one finds $\omega = 2\zeta -1$.

It is generally believed that for all short-range correlated disorder potentials, 
the free energy fluctuations exponent assumes a universal value $\omega = 1/3$.
Numerical studies \cite{numer} as well as the solution via mapping 
to the Burgers equation \cite{hhf_85} confirm this conjecture.
One arrives to the same conclusion studying scaling
properties of the free energy by mapping
the replicated problem to $N$ particle quantum bosons system
\cite{kardar_87} and using the Bethe-Anzats solution. 
However, in this latter case, the resulting distribution function $P_{*}(\tilde{f})$ 
exhibits severe pathologies such as the vanishing of its second 
moment, which assumes that the distribution function is not positively defined.

Let us consider this point in more detail. For the string with the zero boundary conditions
at $\tau=0$ and at $\tau=L$ the partition function of a given sample is
\begin{equation}
\label{bas1-6}
   Z[V] = \int_{\phi(0)=0}^{\phi(L)=0} 
              {\cal D} [\phi(\tau)]  \;  \mbox{\Large e}^{-\beta H[\phi,V]}
\end{equation}
where $\beta = 1/T$ denotes the inverse temperature. On the other hand,
the partition function is related to the total free energy $F[V]$ via
\begin{equation}
\label{bas1-7}
Z[V] = \exp( -\beta  F[V])
\end{equation}
The free energy $F[V]$  is defined for a specific 
realization of the random potential $V$ and thus represent a random variable. 
Let us take the $N$-th power of both sides of Eq.(\ref{bas1-7})
and perform the averaging over the random potential $V$:
\begin{equation}
\label{bas1-8}
Z[N,L] = \overline{\biggl(\exp( -\beta N F[V]) \Biggr)}
\end{equation}
The quantity in the l.h.s of the above equation
\begin{equation}
\label{bas1-9}
Z[N,L] \equiv \overline{Z^{N}[V]} 
\end{equation}
is called the replica partition function, and it is defined originally for an arbitrary {\it integer} parameter $N$. Substituting $F = f_{0} L +  \tilde{f}  L^{\omega}$, 
into Eq.(\ref{bas1-8}) and redefining 
\begin{equation}
\label{bas1-10}
Z[N,L] = \tilde{Z}[N,L] \; \mbox{\Large e}^{-\beta N f_{0} L} 
\end{equation}
we get
\begin{equation}
\label{bas1-11}
\tilde{Z}[N,L] = \overline{\biggl(\exp( -\beta N  L^{\omega} \tilde{f}) \Biggr)}
\end{equation}
The averaging in the r.h.s of the above equation can be represented in terms of the 
distribution function $P_{L}(\tilde{f})$ (which depends on the system size $L$). 
In this way we arrive to the following general relation 
between the replica partition function $\tilde{Z}[N,L]$ and the distribution function 
of the free energy fluctuations $P_{L}(\tilde{f})$:
\begin{equation}
\label{bas1-12}
   \tilde{Z}[N,L] \; =\;
           \int_{-\infty}^{+\infty} d\tilde{f} \, P_{L} (\tilde{f}) \;  
           \mbox{\Large e}^{ -\beta N  L^{\omega} \, \tilde{f}}
\end{equation}
The above equation is the bilateral Laplace transform of  the function $P_{L}(\tilde{f})$,
and at least formally it allows to restore this function in terms of the replica partition function $\tilde{Z}[N,L]$. In order to do so we have to compute $\tilde{Z}[N,L]$ for an {\it arbitrary} 
integer $N$ and then perform  analytical continuation of this function from integer to arbitrary complex values of $N$. 
Introducing  a new {\it complex} variable 
\begin{equation}
\label{bas1-13}
\tilde{s} = \beta N L^{\omega}
\end{equation}
and denoting
\begin{equation}
\label{bas1-14}
\tilde{Z}[\frac{\tilde{s}}{\beta L^{\omega}}, L] \equiv \tilde{Z}_{L}(\tilde{s})
\end{equation}
we could reconstruct the distribution function $P_{L}(\tilde{f})$ 
via the inverse Laplace transform
\begin{equation}
\label{bas1-15}
   P_{L}(\tilde{f}) \; = \;  \int_{-i\infty}^{+i\infty}  \frac{d\tilde{s}}{2\pi i}
       \; \tilde{Z}_{L}(\tilde{s}) \;   \mbox{\Large e}^{\tilde{s} \tilde{f}},
\end{equation}
Finally, provided there are exist a {\it finite} thermodynamic limit function
\begin{equation}
\label{bas1-16}
\lim_{L\to\infty} \tilde{Z}_{L}(\tilde{s}) \equiv \tilde{Z}_{*}(\tilde{s}) 
\end{equation}
we can find the distribution function 
\begin{equation}
\label{bas1-17}
   P_{*}(\tilde{f}) \; = \;  \int_{-i\infty}^{+i\infty}  \frac{d\tilde{s}}{2\pi i}
                \; \tilde{Z}_{*}(\tilde{s}) \;   \mbox{\Large e}^{\tilde{s} \tilde{f}},
\end{equation}
which would describe the statistics of the rescaled free energy fluctuations $\tilde{f}$ 
in the infinite system. The above equation defining
$P_{*}(\tilde{f})$ contains no parameters and hence is expected to be
universal. Therefore according to the relation $\tilde{s} = \beta N L^{\omega}$ we see 
that in the thermodynamic limit the relevant values of the original replica parameter are
\begin{equation}
\label{bas1-18}
N \sim L^{-\omega} \to 0
\end{equation}
This explains why the two limits $L\to\infty$ and $N\to 0$ do not commute \cite{Medina_93},
and the approximation of the replica partition
function through the ground state wave function fails (see also Ref.\
\cite{dirpoly}).
In Kardar's original solution \cite{kardar_87}, after
mapping the replicated problem to interacting quantum bosons, one arrives at
the replica partition function for positive integer parameters $N > 1$.
Assuming a large $L \to \infty$ limit, one is tempted to approximate the
result by the ground state contribution only, as for any $N > 1$ the
contributions of excited states are exponentially small for $L\to\infty$.
However, in the analytic continuation for arbitrary complex $N$ the
contributions which are exponentially small at positive integer $N > 1$ can
become essential in the region $N\to 0$, which according to Eq.\
(\ref{bas1-17}) defines the function $P_{*}(\tilde{f})$
Thus, it is the neglection 
of the excited states which is the origin of non-physical nature 
of the obtained solution. In other words, for the proper analytic continuation 
of the replica solution to the region $N\to 0$, first one has to calculate the replica
partition function $Z[N,L]$ {\it exactly} for arbitrary integer $N$, and only after that
one can take the thermodynamic limit $L\to \infty$ while keeping the value of the
parameter $\tilde{s} = \beta N L^{\omega}$  finite.

In the present paper, we report the results of the calculation of the replica
partition function $Z(N,L)$ for arbitrary integer $N$ which in terms of the
Bethe-Ansatz solution for quantum bosons with attractive $\delta$-interactions
(Sections II - III) involves the summation over the {\it entire spectrum} of
exited states (Section IV).
Unfortunately the analytic continuation of the obtained {\it exact} expression
$Z(N,L)$ from integer to non-integer $N$ 
turns out to be ambiguous, since our replica partition function growth as $\exp(N^3)$
at large $N$ (similar problem of the analytic continuation to the region $N\to 0$
one faces in the replica theory of the mean-field
spin-glasses where the replica partition function growth as $\exp(N^2)$ 
\cite{thanks}).
Performing a kind of a "replica symmetric" analytic continuation, i.e.
just assuming that originally integer-value
parameter $N$ can take arbitrary complex values and
taking the thermodynamic limit $L\to\infty$ in 
 $Z(N,L)$ (Section V) allows us to compute an inverse Laplace
transformation, cf. Eq.\ (\ref{bas1-17}), which provides us with the explicit expression
for the distribution function of the free-energy fluctuations (Section VI,
Eq.\ (\ref{bas6-8})). Although up to the present moment, we have not uncovered 
any unphysical properties in the obtained probability function $P_{*}(f)$,
this solution could be considered as distant analog
of the "replica symmetric approximation" in the mean-field spin-glasses. 
In particular, it should be noted that
our result is {\it different} from the Tracy-Widom distribution \cite{Tracy-Widom},
which describes the statistics of fluctuations in various statistical systems
\cite{PNG_Spohn,LIS,LCS,oriented_boiling,ballistic_decomposition,DP_johansson}
which are widely believed to belong to the same universality class as the
present model \cite{Derrida1,KK,Prahofer-Spohn} (for further discussion of this issue
see Section VII).

Various technical aspects of the
calculations are moved to the Appendices. In particular, in Appendices A, B,
and C, we analyze the structure and properties of $N$-particle wave functions
of one-dimensional quantum bosons, both with repulsive and with attractive
interactions.


\section{Mapping to quantum bosons}

Explicitly, the replica partition function, Eq.(\ref{bas1-9}), of the system described by
the Hamiltonian, Eq.(\ref{bas1-1}), is
\begin{equation}
\label{bas2-1}
   Z(N,L) = \prod_{a=1}^{N} \int_{\phi_{a}(0)=0}^{\phi_{a}(L)=0} 
   {\cal D} \phi_{a}(\tau) \;
   \overline{\exp\Biggl[-\beta \int_{0}^{L} d\tau \sum_{a=1}^{N}
   \bigl\{\frac{1}{2} \bigl[\partial_\tau \phi_{a}(\tau)\bigr]^2 
   + V[\phi_{a}(\tau),\tau]\bigr\}\Biggr] }
\end{equation}
Since it is assumed that the random potential $V[\phi,\tau]$ has the Gaussian 
distribution the disorder average $\overline{(...)}$ in the above equation 
is very simple:
\begin{equation}
\label{bas2-2}
\overline{\exp\Biggl[-\beta \int_{0}^{L} d\tau \sum_{a=1}^{N}
     V[\phi_{a}(\tau),\tau]\Biggr] } \; = \; 
\exp\Biggl[\frac{\beta^{2}}{2} \int\int_{0}^{L} d\tau d\tau' \sum_{a,b=1}^{N}
     \overline{V[\phi_{a}(\tau),\tau] V[\phi_{b}(\tau'),\tau']}\Biggr]
\end{equation}
Using Eq.(\ref{bas1-2}) we have:
\begin{equation}
\label{bas2-3}
   Z(N,L) = \prod_{a=1}^{N} \int_{\phi_{a}(0)=0}^{\phi_{a}(L)=0} 
   {\cal D} \phi_{a}(\tau) \;
   \exp\Biggl[-\frac{1}{2}\beta \int_{0}^{L} d\tau 
   \Bigl\{\sum_{a=1}^{N} \bigl[\partial_\tau \phi_{a}(\tau)\bigr]^2 
   -\beta u \sum_{a,b=1}^{N} \delta\bigl[\phi_{a}(\tau)-\phi_{b}(\tau)\bigr]\Bigr\}\Biggr] 
\end{equation}
It should be noted that the second term in the exponential 
of the above equation contain formally divergent contributions proportional
to $\delta(0)$ (due to the terms with $a=b$). In fact, this is just an indication
that the {\it continuous} model, Eqs.(\ref{bas1-1})-(\ref{bas1-2}) is ill defined
as short distances and requires proper lattice regularization. Of course, the 
corresponding lattice model would contain no divergences, and the terms with
$a=b$ in the exponential of the corresponding replica partition function would produce
irrelevant constant $\frac{1}{2}L\beta^2 u N \delta(0)$ (where the lattice
version of $\delta(0)$ has a finite value). Since the  lattice 
regularization has no impact on the continuous long distance properties 
of the considered system this term will be just omitted in our further study.

Introducing the $N$-component scalar field replica Hamiltonian
\begin{equation}
\label{bas2-4}
   H_{N}[{\boldsymbol \phi}] =  
   \frac{1}{2} \int_{0}^{L} d\tau \Biggl(
   \sum_{a=1}^{N} \bigl[\partial_\tau\phi_{a}(\tau)\bigr]^2 
   - \beta u \sum_{a\not= b}^{N} 
   \delta\bigl[\phi_{a}(\tau)-\phi_{b}(\tau)\bigr] \Biggr)
\end{equation}
for the replica partition function, Eq.(\ref{bas2-3}), 
we obtain the standard expression
\begin{equation}
   \label{bas2-5}
   Z(N,L) = \prod_{a=1}^{N} \int_{\phi_{a}(0)=0}^{\phi_{a}(L)=0} 
   {\cal D} \phi_{a}(\tau) \;
   \mbox{\Large e}^{-\beta H_{N}[{\boldsymbol \phi}] }
\end{equation}
where ${\boldsymbol \phi} \equiv \{\phi_{1},\dots, \phi_{N}\}$.
According to the above definition this partition function describe the statistics
of $N$ $\delta$-interacting (attracting) trajectories $\phi_{a}(\tau)$ all starting 
(at $\tau=0$) and ending (at $\tau=L$) at zero: $\phi_{a}(0) = \phi_{a}(L) = 0$

In order to map the problem to one-dimensional quantum bosons, 
 let us introduce more general object
\begin{equation}
   \label{bas2-6}
   \Psi({\bf x}; t) = 
\prod_{a=1}^{N} \int_{\phi_a(0)=0}^{\phi_a(t)=x_a} {\cal D} \phi_a(\tau)
  \;  \mbox{\Large e}^{-\beta H_{N} [{\boldsymbol \phi}]}
\end{equation}
which describes $N$ trajectories $\phi_{a}(\tau)$ all starting at zero ($\phi_{a}(0) = 0$),
but ending at $\tau = t$ in arbitrary given points $\{x_{1}, ..., x_{N}\}$.
One can easily show that instead of using the path integral, $\Psi({\bf x}; t)$
may be obtained as the solution of the  linear differential equation
\begin{equation}
   \label{bas2-7}
\partial_t \Psi({\bf x}; t) \; = \;
\frac{1}{2\beta}\sum_{a=1}^{N}\partial_{x_a}^2 \Psi({\bf x}; t)
  \; + \; \frac{1}{2}\beta^2 u \sum_{a\not=b}^{N} \delta(x_a-x_b) \Psi({\bf x}; t)
\end{equation}
with the initial condition 
\begin{equation}
   \label{bas2-8}
\Psi({\bf x}; 0) = \Pi_{a=1}^{N} \delta(x_a)
\end{equation}
One can easily see that Eq.(\ref{bas2-7}) is the imaginary-time
Schr\"odinger equation
\begin{equation}
   \label{bas2-9}
-\partial_t \Psi({\bf x}; t) = \hat{H} \Psi({\bf x}; t)
\end{equation}
with the Hamiltonian
\begin{equation}
   \label{bas2-10}
   \hat{H} = 
    -\frac{1}{2\beta}\sum_{a=1}^{N}\partial_{x_a}^2 
   -\frac{1}{2}\beta^2 u \sum_{a\not=b}^{N} \delta(x_a-x_b) 
\end{equation}
which describes  $N$ bose-particles of mass $\beta $ interacting via
the {\it attractive} two-body potential $-\beta^2 u \delta(x)$. 
The original replica partition function, Eq.(\ref{bas2-5}), then is obtained via a particular
choice of the final-point coordinates,
\begin{equation}
   \label{bas2-11}
   Z(N,L) = \Psi({\bf 0};L).
\end{equation}

The standard general strategy of the further calculations is in the following.
Let us denote the eigenfunctions of the Hamiltonian Eq.(\ref{bas2-10}) by $\Psi_{\xi}({\bf x})$
where the index $\xi$ (which can be both integer and continuous) labels the eigenstates.
Provided the wave functions $\Psi_{\xi}({\bf x})$ constitute the orthonormal and complete set, 
the time dependent solution of the equation (\ref{bas2-7}) with the initial conditions, Eq.(\ref{bas2-8}), is given by
\begin{equation}
   \label{bas2-12}
\Psi({\bf x}; t) = \sum_{\xi} \; \Psi_{\xi}({\bf x}) \; \Psi_{\xi}^{*}({\bf 0}) \; 
\mbox{\Large e}^{-E(\xi) t}
\end{equation}
where $E(\xi)$ denotes the energy of the $\xi$-th eigenstate:
\begin{equation}
   \label{bas2-13}
   \hat{H} \Psi_{\xi}({\bf x}) \; = \; E(\xi) \Psi_{\xi}({\bf x})
\end{equation}
Then, according to Eq.(\ref{bas2-11}), the replica partition function $Z(N,L)$ of the original
polymer system is obtained just by the summation over all eigenstates of the quantum Hamiltonian 
(\ref{bas2-10}):
\begin{equation}
   \label{bas2-14}
   Z(N,L) \; = \; \sum_{\xi} \; \big|\Psi_{\xi}({\bf 0}) \big|^{2}\; 
\mbox{\Large e}^{-E(\xi) L}
\end{equation}
Thus, the crucial point of the present approach is finding the eigenfunctions and the 
energy spectrum of the Hamiltonian (\ref{bas2-10}), which is the topic of the next
section.

\section{Eigenstates of the one-dimensional quantum bosons system}

\subsection{Repulsive bosons}

The eigenfunctions  of one-dimensional $\delta$-interacting {\it repulsive} 
($u < 0$) quantum bosons, Eq.(\ref{bas2-10}), have been derived by Lieb and Liniger in 1963 \cite{Lieb-Liniger}.
An eigenstate of this system is characterized by $N$ continuous  momenta
 $\{q_{1},...,q_{N}\} \equiv {\bf q}$ with the wave function (see Appendix A)
\begin{equation}
   \label{bas3-1}
  \Psi_{\bf q}^{(N)}({\bf x}) = C^{(N)}({\bf q})
\sum_{P} (-1)^{[P]} 
\biggl(\prod_{a<b}^{N}\bigl[ (q_{p_a}- q_{p_b}) + i\kappa \sgn(x_{a}-x_{b})\bigr]\biggr) \;
\exp\bigl[ i\sum_{a=1}^{N} q_{p_a} x_{a} \bigr]
\end{equation}
where we have introduced the notation
\begin{equation}
   \label{bas3-2}
\kappa = \beta^{3} u
\end{equation} 
The summation in Eq.(\ref{bas3-1}) goes over all permutations $P$ of the $N$ momenta $\{q_{1},...,q_{N}\}$ over $N$ particles located at $\{x_{1},...,x_{N}\}$, and  $[P]$ denotes 
the parity of the permutation. The normalization constant $C^{(N)}({\bf q})$ is 
\begin{equation}
   \label{bas3-3}
C^{(N)}({\bf q}) = \frac{1}{\sqrt{N! \prod_{a<b}^{N} 
\bigl[ (q_{a}-q_{b})^{2} +\kappa^{2} \bigr] }}
\end{equation}
and the associated energy $E_{N}({\bf q})$ is
\begin{equation}
   \label{bas3-4}
E_{N}({\bf q}) = \frac{1}{2\beta} \sum_{a=1}^{N} q_{a}^{2} 
\end{equation}
A useful alternative  representations  of these wave functions are
\begin{eqnarray}
   \label{bas3-5}
  \Psi_{\bf q}^{(N)}({\bf x}) &=&  C^{(N)}({\bf q})
\sum_{P} (-1)^{[P]} 
\biggl(\prod_{a<b}^{N}\bigl[-i\bigl( \partial_{x_a} - \partial_{x_b}\bigr) +i \kappa \sgn(x_{a}-x_{b})\bigr] \biggr)\;
\exp\bigl[ i\sum_{a=1}^{N} q_{p_a} x_{a} \bigr] \\
\nonumber
\\ 
&=&  C^{(N)}({\bf q})
    \biggl(\prod_{a<b}^{N}\bigl[ -i\bigl(\partial_{x_a} - \partial_{x_b}\bigr) +i \kappa \sgn(x_{a}-x_{b})\bigr]\biggr) \;
    \det\bigl[\widehat{\exp( i {\bf q \, x})} \bigr]
\label{bas3-5a}
\end{eqnarray}
where the symbol $\widehat{\exp( i {\bf q \, x})}$ denotes the $N\times N$ matrix with the 
elements $\exp( i q_{a} x_{b}) \; (a,b = 1,...,N)$.
Here, by definition,  the differential operators $\partial_{x_a}$  act only 
on the exponential terms and {\it not} on the signum functions $\sgn(x_{a}-x_{b})$.
One can easily see that the above wave functions $ \Psi_{\bf q}^{(N)}({\bf x})$ represent
a set of plane waves in any sector of the type $x_{a_1} < x_{a_2} < ... < x_{a_N}$
with a finite jump (equal to $\kappa$) of the derivatives at all "boundary" points 
$x_{a_i} = x_{a_j}$. These functions are symmetric
with respect to any permutation of the particle coordinate $\{x_{1},...,x_{N}\}$ and 
{\it antisymmetric} with respect to permutations of the momenta $\{q_{1},...,q_{N}\}$.
It can be proven that the  wave functions, Eq.(\ref{bas3-1}), are orthonormal and 
constitute the complete set (see Appendix A; a  detailed discussion
one can found e.g. in Refs. \cite{bogolubov,gaudin}). 
Specifically,
for any two functions $\Psi_{\bf q}^{(N)}({\bf x})$ and $\Psi_{\bf q'}^{(N)}({\bf x})$
considered in the sectors $q_{1} < q_{2} < ... < q_{N}$ and   $q'_{1} < q'_{2} < ... < q'_{N}$
orthonormality implies that
\begin{equation}
   \label{bas3-6}
\int_{-\infty}^{+\infty} dx_{1}...dx_{N} \;
{\Psi_{\bf q'}^{(N)}}^{*}({\bf x}) \Psi_{\bf q}^{(N)}({\bf x}) \; = \; 
(2\pi)^{N} \delta(q'_{1}-q_{1}) \delta(q'_{2}-q_{2}) ... \delta(q'_{N}-q_{N})
\end{equation}
Similarly, for any two functions $\Psi_{\bf q}^{(N)}({\bf x})$ and 
$\Psi_{\bf q}^{(N)}({\bf x'})$ considered 
in the sectors $x_{1} < x_{2} < ... < x_{N}$ and   
$x'_{1} < x'_{2} < ... < x'_{N}$, completeness implies that
\begin{equation}
   \label{bas3-7}
\int_{-\infty}^{+\infty} dq_{1}...dq_{N} \; 
\Psi_{\bf q}^{(N)^{*}}({\bf x}) \Psi_{\bf q}^{(N)}({\bf x'}) \; = \; 
(2\pi)^{N} \delta(x_{1}-x'_{1}) \delta(x_{2}-x'_{2}) ... \delta(x_{N}-x'_{N})
\end{equation}
Thus, for a repulsive interaction ($u < 0$) the time dependent solution of the
differential equation (\ref{bas2-7}) (with the starting condition, Eq.(\ref{bas2-8}))
would be sufficiently simple:
\begin{equation}
   \label{bas3-8}
\Psi^{(N)}({\bf x}; t) \; = \; \int_{q_{1}< ... < q_{N}} \; \frac{dq_{1}...dq_{N}}{(2\pi)^{N}} \; 
\Psi_{\bf q}^{(N)^{*}}({\bf x}) \Psi_{\bf q}^{(N)}({\bf 0}) 
\exp\bigl[-\frac{t}{2\beta} \sum_{a=1}^{N} q_{a}^{2} \bigr]
\end{equation}

Unfortunately, from the point of view of the replica theory of disordered polymers the
{\it repulsive} bosons make no physical meaning since the parameter $u$ (according to its
definition, Eq.(\ref{bas1-2})) is positively defined.

\subsection{Attractive bosons}

The situation with {\it attractive} ($u > 0$) bosons is more complicated.
One can easily prove that (irrespective of the sign of the parameter $u$)
the functions $\Psi_{\bf q}^{(N)}({\bf x})$, Eq.(\ref{bas3-1}), are othonormal eigenfunctions
of the Hamiltonian, Eq.(\ref{bas2-10}).
However, unlike the repulsive case, this set of functions is 
{\it not complete}. In other words, for $u > 0$ the completeness conditions, Eqs.(\ref{bas3-7}), 
are not satisfied. Physically this indicates that besides the continuous spectrum (or free particles) states, our system must have another types of eigenstates, in which the particles
are bound into localized clusters.  
The spectrum and some properties of the eigenfunctions
for attractive one-dimensional quantum bosons have been derived by McGuire
\cite{McGuire} and by Yang \cite{Yang} (see also Ref.\ \cite{Takahashi,Calabrese}).
However, since attractive bosons do not have a proper thermodynamic limit
(in the number of particles $N\to\infty$) due
to the scaling $E_N \propto -N^3$, the interest in this system has been rather
limited.

We first consider the ground state  wave function 
$\Psi_{q}^{(1)}({\bf x})$ in which all $N$ particles are bound into one cluster 
with the free center of mass motion controlled by the momentum $q$ (see  Appendix B):
\begin{equation}
   \label{bas3-9}
   \Psi_{q}^{(1)}({\bf x}) \; = \;  C^{(1)}(q) \; 
    \exp\biggl[i q \sum_{a=1}^{N} x_{a} - \frac{1}{4}\kappa \sum_{a\not=b}^{N} |x_{a}-x_{b}| \biggr]
\end{equation}
where
\begin{equation}
   \label{bas3-10}
   C^{(1)}(q) \; = \; \sqrt{\frac{\kappa^N N!}{\kappa N}}
\end{equation}
is the normalization constant defined by the orthonormality condition
\begin{equation}
   \label{bas3-11}
\int_{-\infty}^{+\infty} dx_{1}...dx_{N} \; 
\Psi_{q}^{(1)}({\bf x}) {\Psi_{q'}^{(1)}}^{*}({\bf x}) \; = \; 
(2\pi) \delta(q-q')
\end{equation}
The energy of this state is 
\begin{equation}
   \label{bas3-12}
E_{1}(q;N) = \frac{N}{2\beta} q^2 - \frac{\kappa^{2}}{24\beta}(N^{3}-N)
\end{equation}
On the other hand, one can also easily prove that the above ground state wave function, Eq.(\ref{bas3-9}), can be represented in the form similar to the free particle
structure, Eq(\ref{bas3-1}), by introducing (discrete)
imaginary parts for the momenta $q_{a}$. Indeed, due to the symmetry of this function 
with respect to permutations of $\{x_{1}, x_{2},...,x_{N}\}$ it is sufficient to consider 
it in the sector $x_{1} < x_{2} < ... < x_{N}$. Defining
\begin{equation}
\label{bas3-13}
q_{a} \; = \; q - \frac{i}{2} \kappa (N+1-2a)
\end{equation}
and substituting these momenta into the general expression for the wave function,
Eq.(\ref{bas3-1}) one easily recovers Eq.(\ref{bas3-9}) (see  Appendix B for details)
Also, substituting Eq.(\ref{bas3-13}) into the general expression for the
energy spectrum, Eq.(\ref{bas3-4}) one can also recover, Eq.(\ref{bas3-12}).

Now, using the above scheme, one can  construct 
the eigenfunctions of a generic excited state. It consists of $M$  ($1 \leq M \leq N$)  
"clusters" $\{\Omega_{\alpha}\}$ of  bound particles, where $\alpha = 1,...,M$ labels 
 a given cluster. Each cluster is characterized by the momentum $q_{\alpha}$
of its center of mass motion, and by the number $n_{\alpha}$ of particles contained in it
(such that $\sum_{\alpha=1}^{M} n_{\alpha} = N$).  Instead of $N$ independent real momenta
$q_{a}$ ($a=1,...,N$) one introduces $M$ complex "vector" momenta
\begin{equation}
\label{bas3-16}
q_{a} \; \to \; q^{\alpha}_{r} \; = \; 
q_{\alpha} \; - \; \frac{i}{2} \; \kappa \; (n_{\alpha} + 1 -2 r)
\end{equation}
where $r = 1, 2, ..., n_{\alpha}$ and
\begin{equation}
\label{bas3-17}
\sum_{\alpha = 1}^{M} \; n_{\alpha} \; = \; N
\end{equation}
The corresponding wave function $\Psi_{\bf q, n}^{(M)}(x_1,...,x_N)$ is characterized by 
$M$ continuous parameters ${\bf q} = (q_{1}, ..., q_{M})$ (which are the momenta of the 
center of mass motion of the clusters) and $M$ integer parameters 
${\bf n} = (n_{1}, ..., n_{M})$ (which are the numbers of particles of each cluster).
Explicitly this wave function is given by Eq.(\ref{bas3-5a}) where $N\times N$ matrix
$\widehat{\exp( i {\bf q \, x})}$ is now composed of $N$ columns 
\begin{equation}
   \label{C3a}
    \{ q^{1}_{1}, \; q^{1}_{2},  \; \dots,  \; q^{1}_{n_{1}}; \; 
                    q^{2}_{1}, \; q^{2}_{2}, \; \dots, \;  q^{2}_{n_{2}}; 
                    \; \; \dots \dots \; ;  \; 
                    q^{M}_{1}, \; q^{M}_{2}, \; \dots,  \; q^{M}_{n_{M}} \}
\end{equation}
and $N$ rows $x_{a}$. 
   One can easily see that as in the case of repulsive bosons, 
this wave function is symmetric with respect to permutation of particles coordinates.
However, for practical applications the general representation, Eq.(\ref{bas3-5a}),  
is not very convenient. 
   Writing the determinant of the matrix $\widehat{\exp( i {\bf q \, x})}$ 
explicitly, after a few efforts in
simple algebra one can derive more transparent  structure of the 
wave function  (see Appendix C for details). 
Assuming first that the position of particles are ordered,
 $x_{1} < x_{2} < ... < x_{N}$,
let us  consider a permutation
$P$ of $N$ momenta $q^{\alpha}_{r}$, Eq.(\ref{C3a}),
over $N$ particles $x_{a}$, so that 
a particle number $a$ 
is attributed a momentum component $q^{\alpha(a)}_{r(a)}$.
The particles getting the momenta with the same 
 $\alpha$ (having the same real part $q_{\alpha}$) will be called  
belonging to a cluster $\Omega_{\alpha}$. For a given permutation $P$
the particles belonging 
to the same cluster are  numbered by the  "internal" index $r = 1,...,n_{\alpha}$.
From now on we have to take into account {\it not all the permutations},
but only those for which the internal particle number $r(a)$
is the growing function of the particle number $a$  in every cluster.
Namely, let a cluster $\Omega_{\alpha}$ consists of the particles
$x_{a_{1}}, x_{a_{2}}, \; \dots \; , \; x_{a_{n_{\alpha}}}$
(where $a_{1} < a_{2} < \dots < a_{n_{\alpha}}$ and the positions of particles
are ordered: $x_{a_{1}} < x_{a_{2}} < \; \dots \; < x_{a_{n_{\alpha}}}$). 
Then, among all $n_{\alpha}!$
"internal" permutations of the momenta components 
$q^{\alpha}_{1}, \; q^{\alpha}_{2}, \; \dots \; , \;  q^{\alpha}_{n_{\alpha}}$,
non-zero contribution is given only by the one in which $r(a_{i+1}) = r(a_{i}) +1$.
In this case the explicit form of the wave function (for $M \geq 2$) reeds (see  Appendix C)
\begin{eqnarray}
   \label{bas3-20}
\nonumber
  \Psi_{\bf q, n}^{(M)}({\bf x}) &=& C^{(M)}_{\bf q, n}
{\sum_{P}}' (-1)^{[P]} 
\prod_{\substack{ a<b\\ \alpha(a)\not=\alpha(b) }}^{N}\biggl[ 
\biggl(q_{\alpha(a)} -\frac{i}{2} \kappa (n_{\alpha(a)} + 1 -2 r(a))\biggr) -
\biggl(q_{\alpha(b)} -\frac{i}{2} \kappa (n_{\alpha(b)} + 1 -2 r(b))\biggr)
  - i \kappa \biggr]  \times\\
& &\times\exp\biggl[
i\sum_{a=1}^{N}q_{\alpha(a)} x_{a} 
 +\frac{\kappa}{2}\sum_{a=1}^{N} (n_{\alpha(a)} +1 - 2r(a)) x_{a} \biggr]
\end{eqnarray}
where the product goes only over pairs of particles belonging to {\it different} clusters
and
the symbol ${\sum_{P}}'$ means that the summation goes only over 
the permutations $P$ in which the "internal" indices
$r(a)$ are ordered inside each cluster.

For generic positions of the particles (beyond the sector $x_{1} < x_{2} < ... < x_{N}$)
the expression for the wave function reduces to 
\begin{eqnarray}
\label{C13a}
\nonumber
  \Psi_{\bf q, n}^{(M)}({\bf x}) &=& C^{(M)}_{\bf q, n}
{\sum_{P}}' (-1)^{[P]} 
\prod_{\substack{ a<b\\ \alpha(a)\not=\alpha(b) }}^{N}\biggl[ 
q_{\alpha(a)}- q_{\alpha(b)}
+ \frac{i\kappa}{2} \sum_{c\in\Omega_{\alpha(a)}} \sgn(x_{a}-x_{c}) 
- \frac{i\kappa}{2} \sum_{c\in\Omega_{\alpha(b)}} \sgn(x_{b}-x_{c}) 
+ i\kappa \sgn(x_{a}-x_{b}) \biggr]  
\\
\nonumber
\\
&& \times \exp\biggl[
i\sum_{\alpha=1}^{M} q_{\alpha} \sum_{a\in\Omega_{\alpha}} x_{a} 
 - \frac{\kappa}{4}\sum_{\alpha=1}^{M} \sum_{a,a'\in\Omega_{\alpha}} 
\big|x_{a} - x_{a'}\big| \biggr]
\end{eqnarray}
Here the summation goes only over the permutations in which "internal" indices $r(a)$
in the clusters are ordered according to the spatial ordering of the particles belonging to these
clusters. For example, let a cluster $\Omega_{\alpha}$ be composed of the particles 
$\{ x_{a_{1}}, x_{a_{2}}, \; \dots \; , \; x_{a_{n_{\alpha}}} \}$ and the spatial 
positions of these particles are such that 
$x_{a_{1}} < x_{a_{2}} < \; \dots \; < x_{a_{n_{\alpha}}}$, while the particles numbering
$a_{1}, a_{2}, \dots, a_{n_{\alpha}}$ is now {\it arbitrary}. Then the ordering of the internal
index $r(a)$ of the permutations involved in  Eq.(\ref{C13a}) is such that
$r(a_{i+1}) = r(a_{i}) +1$ (i.e. it goes from the smallest $x_{a}$ in the cluster
to the largest one).

One can easily see that the expression for $\Psi_{\bf q, n}^{(M)}({\bf x})$,
Eq.(\ref{C13a}), can also be rewritten in more compact form:
\begin{equation}
 \label{bas3-19}
 \Psi_{\bf q, n}^{(M)}({\bf x}) =  C^{(M)}_{\bf q, n}
{\sum_{P}}' (-1)^{[P]} 
\prod_{\substack{ a<b\\ \alpha(a)\not=\alpha(b) }}^{N}
\biggl[ -i\bigl(\partial_{x_a} - \partial_{x_b}\bigr) 
+ i \kappa \sgn(x_{a}-x_{b})]\biggr]
\exp\biggl[
i\sum_{\alpha=1}^{M}q_{\alpha}\sum_{a\in\Omega_{\alpha}}x_{a} 
 -\frac{\kappa}{4}\sum_{\alpha=1}^{M}\sum_{a,a'\in\Omega_{\alpha}} |x_{a}-x_{a'}| \biggr]
\end{equation}

\vspace{5mm}

In order to describe the orthogonality  of these wave functions, we have to specify their 
symmetry structure. First of all, like in the case of repulsion, Eq.(\ref{bas3-1}),
the wave function of unbound particles (the case $M=N$ and $n_{1} = n_{2} = ... = n_{N} =1$) 
is fully antisymmetric with respect to any momenta permutation. 
For the generic case, $2\leq M < N$ the situation is slightly more tricky
as the symmetry with respect to the momenta permutations depends on the values
of the corresponding integer parameters $n_{\alpha}$. According to Eq.(\ref{bas3-5a}), 
the permutation of any two momenta $q_{\alpha_{1}}$ and $q_{\alpha_{2}}$ belonging to 
the clusters which have {\it the same} numbers of particles, 
$n_{\alpha_{1}}=n_{\alpha_{2}} = n$
produces the factor $(-1)^{n}$  (this operation corresponds to the permutation of
$n$ columns of the matrix $\exp(i q_{a} x_{b})$). Hence two eigenstates which differ
one from another only by such momenta permutations could be called {\it equivalent}.
 On the other hand, the permutation
of the momenta between two clusters with {\it different} numbers of particles reveals
no specific symmetry at all. In other words, the wave functions with permuted momenta
belonging to two clusters with different numbers of particles are just two different
wave functions describing two {\it different} eigenstates. 
Thus for comparing the wave functions 
described by the parameters $(q_{\alpha}, n_{\alpha}) \;\; (\alpha = 1, ..., M)$
(in particular for the study of their orthogonality) 
it is crucial to specify "subsets" of equal $n$'s. Namely, a generic eigenstate
$({\bf q}, {\bf n})$ with $M$ clusters could be specified in terms of the 
following set of parameters:
\begin{equation}
\label{bas3-21}
({\bf q}, {\bf n}) \; = \;  \{
(\underbrace{q_{1}, m_{1}), ..., (q_{s_{1}}, m_{1})}_{s_{1}}; 
 \underbrace{(q_{s_{1}+1}, m_{2}), ..., (q_{s_{1}+s_{2}}, m_{2})}_{s_{2}}; 
\; .... \; ; 
 \underbrace{(q_{s_{1}+...+s_{k-1}+1}, m_{k}), ..., (q_{s_{1}+...+s_{k}}, m_{k})}_{s_{k}}\}
\end{equation}
where  $s_{i} \; (i=1,...k ; \; 1 \leq k \leq M)$ are the numbers of clusters
 which have the same numbers of particles and $k$ denotes the number of different 
cluster types. For a given $k$
\begin{equation}
\label{bas3-22}
s_{1} + s_{2} + ... + s_{k} \; = \; M
\end{equation}
and 
\begin{equation}
\label{bas3-23}
\sum_{\alpha=1}^{M} n_{\alpha} \; = \; \sum_{i=1}^{k} s_{i} m_{i} \; =  \; N
\end{equation}
In this representations all the integers $\{m_{i}\}$ are assumed to be different:
\begin{equation}
\label{bas3-24}
1 \leq m_{1} < m_{2} < ... < m_{k}
\end{equation}
Due to the symmetry  with respect to the momenta permutations inside the subsets 
of equal $n$'s it is sufficient to consider the wave functions in the sectors
\begin{eqnarray}
\label{bas3-25}
&&
q_{1} < q_{2} < ... < q_{s_{1}} \; ;\\
\nonumber
&&
q_{s_{1}+1} < q_{s_{1}+2} < ... < q_{s_{1}+s_{2}} \; ; \\
\nonumber
&&
................. \\
\nonumber
&&
q_{s_{1}+...+s_{k-1}+1} < q_{s_{1}+...+s_{k-1}+2} < ... < q_{s_{1}+...+s_{k-1}+s_{k}}
\end{eqnarray}
It can be shown (see Appendix C) that any two wave functions 
$\Psi_{\bf q, n}^{(M)}({\bf x})$ and $\Psi_{\bf q', n'}^{(M')}({\bf x})$
in which the parameters $({\bf q, n})$ and $({\bf q', n'})$ are assumed to have
the structure described above  (Eqs.(\ref{bas3-21})-(\ref{bas3-25})), 
are orthogonal:
\begin{equation}
\label{bas3-26}
\int_{-\infty}^{+\infty} dx_{1}...dx_{N} \;
\Psi_{\bf q, n}^{(M)}({\bf x}) \; \Psi_{\bf q', n'}^{(M')^{*}}({\bf x})  \; = \;
{\boldsymbol \delta}(M,M') \; 
\biggl(\prod_{\alpha=1}^{M} 
{\boldsymbol \delta}(n_{\alpha},n'_{\alpha}) \biggr)
\biggl(\prod_{\alpha=1}^{M} (2\pi) \delta(q_{\alpha}-q'_{\alpha}) \biggr)
\end{equation}
where ${\boldsymbol \delta}(n,m)$ is the Kronecker symbol and $\delta(q)$ is the 
$\delta$-function. The above orthonormality condition defines the 
normalization constant
\begin{equation}
   \label{bas3-27}
   C^{(M)}({\bf q,n}) = \Biggl[\frac{1}
   {N! \; \prod_{\alpha<\beta}^{M} \prod_{r=1}^{n_{\alpha}} 
   \prod_{r'=1}^{n_{\beta}}
   \bigl[ [q_{\alpha}-q_{\beta}- i\kappa (\frac{n_{\alpha}-n_{\beta}}{2} 
   - r + r' )]^{2}+\kappa^{2} \bigr]} \; 
   \prod_{\alpha=1}^{M} \frac{(n_{\alpha}!)^{2}\kappa^{n_{\alpha}} }
   {n_{\alpha} \kappa}\Biggr]^{1/2}.
\end{equation}
In other words, the wave functions, Eqs.(\ref{C13a}) or (\ref{bas3-19}) 
form the orthonormal set. Although, at present we are not able to prove that
this set is complete, the {\it suggestion} of the completeness 
(which assumes that there are exist no other eigenstates besides those described above)
looks quite natural.

Finally, substituting Eq.(\ref{bas3-16})-(\ref{bas3-17}) into Eq.(\ref{bas3-4}), 
for the energy spectrum one easily obtains:
\begin{equation}
\label{bas3-18}
E_{M}({\bf q,n}) \; = \;
\frac{1}{2\beta} \sum_{\alpha=1}^{M} \; \sum_{r=1}^{n_{\alpha}} (q^{\alpha}_{k})^{2} 
\; = \; \frac{1}{2\beta} \sum_{\alpha=1}^{M} \; n_{\alpha} q_{\alpha}^{2} \;
- \; \frac{\kappa^{2}}{24\beta}\sum_{\alpha=1}^{M} (n_{\alpha}^{3}-n_{\alpha})
\end{equation}

\section{Replica partition function}

The time dependent solution $ \Psi({\bf x},t)$ of Eq.(\ref{bas2-7}), 
satisfying the starting condition Eq.(\ref{bas2-8}) can be represented in terms of the 
linear combination of the eigenfunctions $\Psi_{\bf q, n}^{(M)}({\bf x})$, 
Eq.(\ref{bas3-20})-(\ref{bas3-19}):
\begin{equation}
\label{bas4-1}
\Psi({\bf x},t) \; = \; \sum_{M=1}^{N} \;
{\sum_{{\bf n}}}' \; 
\int ' {\cal D} {\bf q} \; \; 
 \Psi_{\bf q, n}^{(M)}({\bf x}) \Psi_{\bf q, n}^{(M)^{*}}({\bf 0}) \; 
\exp\bigl[-E_{M}({\bf q,n}) \; t \bigr]
\end{equation}
Here the summations over $n_{\alpha}$ are performed in terms of the parameters $\{s_{i}, m_{i}\}$,
Eqs.(\ref{bas3-22})-(\ref{bas3-24}):
\begin{equation}
\label{bas4-2}
{\sum_{{\bf n}}}' \; \equiv \;
\sum_{k=1}^{M} \; \; 
\sum_{s_{1}...s_{k}=1}^{\infty} \; \; 
\sum_{1 \leq m_{1} ... < m_{k}}^{\infty}
\; {\boldsymbol \delta}\biggl(\sum_{i=1}^{k} s_{i}, \; M\biggr)
\; \;{\boldsymbol \delta}\biggl(\sum_{i=1}^{k} s_{i} m_{i}, \; N\biggr)
\end{equation}
where for simplicity, due to the presence of the Kronecker symbols  the summations over $m_{i}$ and $s_{i}$
are extended to infinity. The symbol $\int '{\cal D} {\bf q}$ in Eq.(\ref{bas4-1}) denotes the integration 
over $M$ momenta $q_{\alpha}$ in the sectors, Eq.(\ref{bas3-25}); 
the energy spectrum $E_{M}({\bf q,n})$ is given by Eq.(\ref{bas3-18})

Now according Eq.(\ref{bas2-11}) for the replica partition function of the original 
directed polymer problem we get
\begin{equation}
\label{bas4-3}
Z(N,L) \; = \; \Psi({\bf 0};L) \; = \; 
\sum_{M=1}^{\infty} \;
{\sum_{{\bf n}}}' \; 
\int ' {\cal D} {\bf q} \; \; 
\bigl| \Psi_{\bf q, n}^{(M)}({\bf 0}) \bigr|^2 \; 
\exp\bigl[-E_{M}({\bf q,n}) \; L \bigr]
\end{equation}
where  due to the presence of the Kronecker symbols  in Eq.(\ref{bas4-2})
the summation over $M$ can also be extended to infinity. 
Using Eq.(\ref{bas3-20}), and taking into account antisymmetry with respect to the momenta permutations,
one can easily prove that for $M \geq 2$,
\begin{eqnarray}
\nonumber
\Psi_{\bf q, n}^{(M)}({\bf 0}) 
&=& 
C^{(M)}_{\bf q, n}
{\sum_{P}}' (-1)^{[P]} 
\prod_{\substack{ a<b\\ \alpha(a)\not=\alpha(b) }}^{N}\biggl( q_{\alpha(a)}- q_{\alpha(b)}
- i \kappa \bigl[\frac{n_{\alpha(a)}-n_{\alpha(b)}}{2}-r(a)+r(b)\bigr]  -i \kappa \biggr) 
\\
\nonumber
\\
&=& 
C^{(M)}_{\bf q, n}
{\sum_{P}}' (-1)^{[P]} 
\prod_{\substack{ a<b\\ \alpha(a)\not=\alpha(b) }}^{N}\biggl( q_{\alpha(a)}- q_{\alpha(b)}
- i \kappa \bigl[\frac{n_{\alpha(a)}-n_{\alpha(b)}}{2}-r(a)+r(b)\bigr]\biggr) 
\label{bas4-4}
\end{eqnarray}
Given the
antisymmetry of the product with respect to permutations of the momenta, it is
sufficient to consider only one (trivial) permutation and multiply the result
by the total number of permutations:
\begin{equation}
\label{bas4-5}
\Psi_{\bf q, n}^{(M)}({\bf 0}) \; = \; 
C^{(M)}_{\bf q, n} \; \frac{N!}{n_{1}! n_{2}! ... n_{M}!} \;
\prod_{\alpha<\beta}^{M} \prod_{r=1}^{n_{\alpha}} \prod_{r'=1}^{n_{\beta}}
\biggl( q_{\alpha} - q_{\beta} 
-i \kappa \bigl[\frac{n_{\alpha}-n_{\beta}}{2} - r + r'\bigr]\biggr)
\end{equation}
For $M=1$, according to Eqs.(\ref{bas3-9})-(\ref{bas3-10}),
\begin{equation}
\label{bas4-6}
\Psi_{q}^{(1)}({\bf 0}) \; = \;  \sqrt{\frac{\kappa^N N!}{\kappa N}}
\end{equation}
We see that the expression $\bigl| \Psi_{\bf q, n}^{(M)}({\bf 0}) \bigr|^2 \; 
\exp\bigl[-E_{M}({\bf q,n}) \; L \bigr]$ in Eq.(\ref{bas4-3}) is symmetric 
with respect to the permutations of the momenta $q_{\alpha}$ belonging 
to the clusters with the same number of particles.
In this case the expression for the partition function can be written in
terms of the {\it unconstrained} integration over the momenta $q_{\alpha}$:
\begin{equation}
\label{bas4-7}
Z(N,L) \; = \;  
\sum_{M=1}^{\infty} \; 
\sum_{k=1}^{M}
\sum_{s_{1}...s_{k}=1}^{\infty} \; \; 
\sum_{1 \leq m_{1} ... < m_{k}}^{\infty} \; 
\frac{
{\boldsymbol \delta}\bigl(\sum_{i} s_{i}, \; M\bigr) \;
{\boldsymbol \delta}\bigl(\sum_{i} s_{i} m_{i}, \; N\bigr) }{
s_{1}! s_{2}! ... s_{k}!}
\biggl(\prod_{\alpha=1}^{M} \int_{-\infty}^{+\infty} \; \frac{dq_{\alpha}}{2\pi} \biggr) \;
\bigl| \Psi_{\bf q, n}^{(M)}({\bf 0}) \bigr|^2 \; 
\mbox{\Large e}^{-E_{M}({\bf q,n}) L}
\end{equation}
where
\begin{equation}
 \label{bas4-8a}
{\bf n} \; \equiv \;   \{\underbrace{m_{1}, ..., m_{1}}_{s_{1}}, 
                         \underbrace{m_{2}, ..., m_{2}}_{s_{2}}, ... ..., 
                         \underbrace{m_{k}, ..., m_{k}}_{s_{k}} \}
\end{equation}
Eq.(\ref{bas4-7}) contains the summations of the quantity
\begin{equation}
\label{bas4-8}
f(n_{1}, n_{2}, ..., n_{M}) \; = \; 
\biggl(\prod_{\alpha=1}^{M} \int_{-\infty}^{+\infty} \; \frac{dq_{\alpha}}{2\pi} \biggr) \;
\bigl| \Psi_{\bf q, n}^{(M)}({\bf 0}) \bigr|^2 \; 
\mbox{\LARGE e}^{-E_{M}({\bf q,n}) L}
\end{equation} 
which is the function of $M$ integer parameters $n_{\alpha}$.
Using explicit expressions, Eq.(\ref{bas3-18}), (\ref{bas3-27}) and 
(\ref{bas4-5}) one can easily prove that this function 
is fully symmetric 
with respect to permutations of {\it all} its $M$ arguments.
In this case 
\begin{equation}
 \label{bas4-8b}
\sum_{k=1}^{M}
\sum_{s_{1}...s_{k}=1}^{\infty} \; \; 
\sum_{1 \leq m_{1} ... < m_{k}}^{\infty} \; 
\frac{
{\boldsymbol \delta}\bigl(\sum_{i} s_{i}, \; M\bigr) \;
{\boldsymbol \delta}\bigl(\sum_{i} s_{i} m_{i}, \; N\bigr) }{
s_{1}! s_{2}! ... s_{k}!} \; f({\bf n}) \; = \; 
\frac{1}{M!} 
\sum_{n_{1}=1}^{\infty}
...  
\sum_{n_{M}=1}^{\infty} 
{\boldsymbol \delta}\biggl(\sum_{\alpha=1}^{M} n_{\alpha}, \; N \biggr) \;
 f({\bf n})
\end{equation}
so that the summations 
in  the expression for the replica partition 
function, Eq.(\ref{bas4-7}), can be essentially simplified:
\begin{equation}
\label{bas4-9}
Z(N,L) \; = \;  
\sum_{M=1}^{\infty} \; 
\frac{1}{M!}
\biggl(\prod_{\alpha=1}^{M} \int_{-\infty}^{+\infty} \; \frac{dq_{\alpha}}{2\pi} \biggr) \;
\sum_{n_{1}=1}^{\infty}
\sum_{n_{2}=1}^{\infty}
\; ... \; 
\sum_{n_{M}=1}^{\infty} \;
{\boldsymbol \delta}\biggl(\sum_{\alpha=1}^{M} n_{\alpha}, \; N \biggr) \;
\bigl| \Psi_{\bf q, n}^{(M)}({\bf 0}) \bigr|^2 \; 
\mbox{\LARGE e}^{-E_{M}({\bf q,n}) L}
\end{equation}
Substituting here the explicit expressions 
for $\Psi_{\bf q, n}^{(M)}({\bf 0})$, Eqs(\ref{bas4-5})-(\ref{bas4-6}), 
for $E_{M}({\bf q,n})$, Eqs.(\ref{bas3-18}), as well as for the normalization constant $C^{(M)}({\bf q,n})$, Eq.(\ref{bas3-27}), 
one gets
\begin{eqnarray}
\nonumber
Z(N.L) &=&
N! \; \kappa^{N} 
 \int_{-\infty}^{+\infty} \frac{dq}{2\pi\kappa N} \;
\mbox{\LARGE e}^{-\frac{N L}{2\beta} q^{2} + \frac{\kappa^{2}L}{24\beta} (N^{3} -N)} \; +
\\
\nonumber
&+& N! \; \kappa^{N} 
\sum_{M=2}^{\infty} \frac{1}{M!} \;
\biggl[
\prod_{\alpha=1}^{M}
\sum_{n_{\alpha}=1}^{\infty}
\int_{-\infty}^{+\infty} \frac{d q_{\alpha}}{2\pi\kappa n_{\alpha}} 
\biggr]
\;{\boldsymbol \delta}\biggl(\sum_{\alpha=1}^{M} n_{\alpha}, \; N \biggr) \; 
\mbox{\LARGE e}^{-\frac{L}{2\beta}\sum_{\alpha=1}^{M} n_{\alpha} q_{\alpha}^{2} + 
\frac{\kappa^{2}L}{24\beta} \sum_{\alpha=1}^{M} (n_{\alpha}^{3} - n_{\alpha})} \; \times
\\
&\times&
\prod_{\alpha<\beta}^{M} \prod_{r=1}^{n_{\alpha}} \prod_{r'=1}^{n_{\beta}}
\frac{\big|q_{\alpha}-q_{\beta}-\frac{i\kappa}{2}(n_{\alpha}-n_{\beta}-2r+2r')\big|^{2}}{
     \bigl[q_{\alpha}-q_{\beta}-\frac{i\kappa}{2}(n_{\alpha}-n_{\beta}-2r+2r')\bigr]^{2} + \kappa^{2}} 
\label{bas4-11}
\end{eqnarray}
The first term in the above expression is the contribution of the ground state $(M=1)$,
and the next terms $(M \geq 2)$ are the contributions of the rest of the energy
spectrum. Next, 
after a few lines of slightly cumbersome transformations (see Appendix D)
Eq.(\ref{bas4-11}) can be reduced to the form (cf. Eq.(\ref{bas1-10}))
\begin{equation}
   \label{bas4-12}
 Z(N,L) \; = \; N! \; \kappa^{N} \; \mbox{\LARGE e}^{-\beta N L f_{0}} \;\;
\tilde{Z}(N,L) 
\end{equation}
where $f_{0}$ the linear (selfaveraging) free energy density,
\begin{equation}
\label{bas4-15}
f_{0} \; = \; \frac{1}{24}\beta^4 u^2 \; ,
\end{equation}
and 
\begin{eqnarray}
\nonumber
\tilde{Z}(N,\lambda) &=& 
 \int_{-\infty}^{+\infty} \frac{dp}{4\pi} \int_{0}^{+\infty} dt \; \;
\mbox{\LARGE e}^{ -\lambda N t - \lambda N p^{2} + \frac{1}{3} \lambda^{3} N^{3}} \; +\\
\nonumber
\\
\nonumber
 &+& \sum_{M=2}^{\infty} \frac{1}{M!} \; \sum_{n_{1}...n_{M}=1}^{\infty} 
\biggl[\prod_{\alpha=1}^{M} 
\int_{-\infty}^{+\infty} \frac{dp_{\alpha}}{4\pi} \int_{0}^{+\infty} dt_{\alpha} \; \; 
\mbox{\LARGE e}^{-\lambda n_{\alpha} t_{\alpha}-\lambda n_{\alpha} p_{\alpha}^{2} 
+ \frac{1}{3} \lambda^{3} n_{\alpha}^{3}} \biggr]
\;{\boldsymbol \delta}\biggl(\sum_{\alpha=1}^{M} n_{\alpha}, \; N \biggr) \; \times \\
&\times&
\prod_{\alpha<\beta}^{M}
\frac{\big|p_{\alpha}-p_{\beta} -i\lambda(n_{\alpha}-n_{\beta})\big|^{2}}{
      \big|p_{\alpha}-p_{\beta} -i\lambda(n_{\alpha}+n_{\beta})\big|^{2}}
\label{bas4-13}
\end{eqnarray}
Here instead of the system length $L$ we have introduced a new parameter
\begin{equation}
   \label{bas4-14}
\lambda (L) \; = \; \frac{1}{2} \biggl(\frac{L}{\beta} \kappa^{2} \biggr)^{1/3} \; = \; 
\frac{1}{2} \biggl(\beta^{5} u^{2} L \biggr)^{1/3}
\end{equation}
Next, we linearize the terms cubic in
$n_{\alpha}$ in the exponentials of Eq.\ (\ref{bas4-13}) with the help of Airy
functions, using the standard relation 
\begin{equation}
   \label{bas4-16}
\exp\bigl( \frac{1}{3} \lambda^{3} n^{3}\bigr) \; = \; 
\int_{-\infty}^{+\infty} dy \; \Ai(y) \; \exp(\lambda n)
\end{equation}
After shifting the Airy function parameters of integration
$y_{\alpha} \to y_{\alpha} + t_{\alpha} + p_{\alpha}^{2}$ the expression 
for $\tilde{Z}(N,\lambda)$ becomes sufficiently compact:
\begin{eqnarray}
\nonumber
\tilde{Z}(N,\lambda) &=& 
\int_{-\infty}^{+\infty} dy \int_{-\infty}^{+\infty} \frac{dp}{4\pi} \int_{0}^{+\infty} dt
\; \; \Ai(y+t+p^{2}) \; \exp(\lambda N y) \; + \\
\nonumber
\\
\nonumber
 &+& \sum_{M=2}^{\infty} \frac{1}{M!} \;
\biggl[\prod_{\alpha=1}^{M} \int_{-\infty}^{+\infty} dy_{\alpha}
\int_{-\infty}^{+\infty} \frac{dp_{\alpha}}{4\pi} \int_{0}^{+\infty} dt_{\alpha} \; 
\Ai(y_{\alpha}+t_{\alpha}+p^{2}_{\alpha}) \biggr] \times \\
&\times& 
\sum_{n_{1}...n_{M}=1}^{\infty} 
{\boldsymbol \delta}\biggl(\sum_{\alpha=1}^{M}n_{\alpha}, \; N\biggr)
\prod_{\alpha<\beta}^{M}
\frac{\big|p_{\alpha}-p_{\beta} -i\lambda(n_{\alpha}-n_{\beta})\big|^{2}}{
      \big|p_{\alpha}-p_{\beta} -i\lambda(n_{\alpha}+n_{\beta})\big|^{2}} 
\prod_{\alpha=1}^{M}
\exp(\lambda n_{\alpha} y_{\alpha})
\label{bas4-17}
\end{eqnarray}
Finally, after performing summations over $\{n_{1}, ..., n_{M}\}$ (see  Appendix E) 
the above expression 
can be represented as an analytic function of two parameters: $\lambda N$ and $\lambda$:
\begin{equation}
   \label{bas4-18}
\tilde{Z}(N,\lambda) \; = \; Z_{1}(\lambda N) 
\; + \; \sum_{M=2}^{\infty} Z_{M}(\lambda N; \lambda)
\end{equation}
where
\begin{equation}
   \label{bas4-19}
Z_{1}(\lambda N) \; = \; 
\int_{-\infty}^{+\infty} dy \int_{-\infty}^{+\infty} \frac{dp}{4\pi} \int_{0}^{+\infty} dt \;
\Ai(y+t+p^{2}) \; \exp(\lambda N y)
\end{equation}
and
\begin{equation}
\label{bas4-20}
Z_{M}(\lambda N; \lambda) \; = \; \frac{1}{(M-1)!}\; \int {\cal D}_{M}({\bf y,p})
\; \int \hat {\cal G}_{M}({\bf p}; \; {\boldsymbol \psi}, {\boldsymbol \phi}, {\boldsymbol \chi})
\; \; \mbox{\LARGE e}^{\lambda N (y_{1}+ i\eta_{1})} \; 
\prod_{\alpha=2}^{M} \frac{\mbox{\LARGE e}^{\lambda (y_{\alpha}+ i\eta_{\alpha})}}{
\mbox{\LARGE e}^{\lambda (y_{1}+ i\eta_{1})} - 
\mbox{\LARGE e}^{\lambda (y_{\alpha}+ i\eta_{\alpha})}}
\end{equation}
with the definition
\begin{equation}
   \label{bas4-22}
\eta_{\alpha} \; \equiv \; 
\eta_{\alpha}({\boldsymbol \psi}, {\boldsymbol \phi}, {\boldsymbol \chi}) 
\; = \; \frac{1}{2} \sum_{\beta\not=\alpha}^{M}
\biggl(\psi_{\alpha\beta}^{2} + \psi_{\beta\alpha}^{2} - \phi_{\alpha\beta}^{2} - \phi_{\beta\alpha}^{2}
+2 \chi_{\alpha\beta} -2 \chi_{\beta\alpha} \biggr)
\end{equation}
Above we have introduced  the integration operator
\begin{equation}
   \label{bas4-21}
\int {\cal D}_{M}({\bf y,p}) \; \equiv \;
\prod_{\alpha=1}^{M} \int_{-\infty}^{+\infty} dy_{\alpha}
\int_{-\infty}^{+\infty} \frac{dp_{\alpha}}{4\pi} \int_{0}^{+\infty} dt_{\alpha} 
\Ai(y_{\alpha}+t_{\alpha}+p^{2}_{\alpha}) 
\end{equation}
as well as the integro-differential operator
\begin{equation}
\label{bas4-23}
\int \hat {\cal G}_{M}({\bf p}; \; {\boldsymbol \psi}, {\boldsymbol \phi}, {\boldsymbol \chi}) 
\; \equiv \; 
\biggl[\prod_{\alpha\not=\beta}^{M} 
\int\int_{-\infty}^{+\infty} \frac{d\psi_{\alpha\beta}d\phi_{\alpha\beta}}{2\pi} 
\frac{\partial}{\partial\chi_{\alpha\beta}} \biggr]\; 
\mbox{\LARGE e}^{-\frac{1}{2}\sum_{\alpha\not=\beta}^{M}\big|p_{\alpha}-p_{\beta}\big|
(\psi_{\alpha\beta}^{2}+\phi_{\alpha\beta}^{2} - 2\chi_{\alpha\beta}) }
\end{equation}
where it is assumed that the derivatives over $\{\chi_{\alpha\beta}\}$ are taken at $\chi_{\alpha\beta}=0$.

The crucial point is that all these factors: $\int {\cal D}_{M}({\bf y,p})$, 
$\int \hat {\cal G}_{M}({\bf p}; \; {\boldsymbol \psi}, {\boldsymbol \phi}, {\boldsymbol \chi}) $,
$\eta_{\alpha}({\boldsymbol \psi}, {\boldsymbol \phi}, {\boldsymbol \chi}) $
as well as the last product in Eq.(\ref{bas4-20}) do not contain the replica
parameter $N$; the latter enters the expression only in the combination $\lambda N$ 
within the exponentials of Eqs.(\ref{bas4-19})-(\ref{bas4-20}).
Thus, we have obtained the exact expression for the partition function, 
Eqs.(\ref{bas4-18})-(\ref{bas4-20}), in the form of an analytic function of the 
replica parameter $N$, which until now was assumed to be an {\it arbitrary integer}. 
In the following, we will consider the
analytic continuation of this function to arbitrary complex values of $N$ and,
in particular, in the limit $N \to 0$. Unfortunately this crucial step of the
analytic continuation is ambiguous, as our partition 
function grows as $\exp(N^3)$ at large $N$ (it is well known that in this case
there can exist many distribution functions which have the same moments).
We will return to this problem with a further discussion in Section VII.

\section{Thermodynamic limit}

Assuming now that the parameter $N$ is an arbitrary complex quantity we are going to take
the thermodynamic limit $L \to \infty$ in the replica partition function, Eqs.(\ref{bas4-18})-(\ref{bas4-20}),
where, according to Eq.(\ref{bas4-14}), the length $L$ enters in terms of the
parameter $\lambda(L) = \frac{1}{2} (\beta^{5} u^{2} L )^{1/3}$. 
It is crucial that the replica parameter
$N$ appears in Eqs.(\ref{bas4-18})-(\ref{bas4-20}) only in the combination $\lambda N$.
Keeping in mind further calculations of the free energy distribution function,
Eqs.(\ref{bas1-12})-(\ref{bas1-18}), in the limit $\lambda \to \infty$ we have to keep the value of the parameter 
\begin{equation}
   \label{bas5-1}
s \; \equiv \;  \lambda N \; = \; \frac{1}{2} (\beta^{5} u^{2} L )^{1/3} \; N
\end{equation}
finite. In other words, in the thermodynamic limit $L\to\infty$, 
the replica parameter $N \sim 1/\lambda \sim L^{-1/3} \to 0$.

It turns out that in the thermodynamic limit the expression 
for the replica partition function, 
Eqs.(\ref{bas4-18})-(\ref{bas4-20}), simplifies dramatically. Indeed, since the functions $\eta_{\alpha}$, Eq.(\ref{bas4-22}), take only {\it real} values, we have
\begin{equation}
   \label{bas5-2}
\lim_{\lambda\to\infty} \frac{\exp\biggl[\lambda (y_{\alpha}+ i\eta_{\alpha})\biggr]}{
\exp\biggl[\lambda (y_{1}+ i\eta_{1})\biggr] - \exp\biggl[\lambda (y_{\alpha}+ i\eta_{\alpha})\biggr]} \; = \; 
\left\{
\begin{array}{lll}
0  \; , \; &\mbox{for}& \; y_{\alpha} < y_{1}\\
\\
-1 \; , \; &\mbox{for}& \;  y_{\alpha} > y_{1} \\
\end{array}
\right.
\end{equation}
Substituting this into Eq.(\ref{bas4-20}) we get
\begin{equation}
   \label{bas5-3}
\lim_{\lambda\to\infty} Z_{M}(s; \lambda) \; \equiv \tilde{Z}_{M}(s)
\; = \; \frac{(-1)^{M-1}}{(M-1)!} 
\int {\cal D}_{M}({\bf y,p})  \; \; \mbox{\LARGE e}^{ s y_{1}} \; 
\biggl[\prod_{\alpha=2}^{M} \theta(y_{\alpha} - y_{1}) \biggr]
  \; \int \hat {\cal G}_{M}({\bf p}; \; {\boldsymbol \psi}, {\boldsymbol \phi}, {\boldsymbol \chi}) \; 
\mbox{\LARGE e}^{i s  \eta_{1}({\boldsymbol \psi}, {\boldsymbol \phi}, {\boldsymbol \chi})}
\end{equation}
It can be shown (see Appendix F) that the last term in the above equation is unity
\begin{equation}
   \label{bas5-6}
\int \hat {\cal G}_{M}({\bf p}; \; {\boldsymbol \psi}, {\boldsymbol \phi}, {\boldsymbol \chi}) \; 
\mbox{\LARGE e}^{i s  \eta_{1}({\boldsymbol \psi}, {\boldsymbol \phi}, {\boldsymbol \chi})} 
\; \equiv \; 1
\end{equation}
Thus, introducing the function
\begin{equation}
\label{bas5-7}
\Phi(x) \; \equiv \; \int_{-\infty}^{+\infty} \frac{dp}{4\pi} \int_{0}^{+\infty} dt \;
\Ai(x + t + p^{2})
\end{equation}
and substituting the definition of the operator $\int {\cal D}_{M}({\bf y,p})$, Eq.(\ref{bas4-21}), into Eq.(\ref{bas5-3}) one gets 
\begin{equation}
\label{bas5-8}
\tilde{Z}_{M}(s) \; = \; \frac{(-1)^{M-1}}{(M-1)!} 
\int_{-\infty}^{+\infty} dx \; \Phi(x) \;  \mbox{\LARGE e}^{s x} \;
\biggl[\int_{x}^{+\infty} dy  \Phi(y) \biggr]^{M-1} 
\end{equation}
Now, substituting Eqs.(\ref{bas5-8}) and (\ref{bas4-19}) into Eq.(\ref{bas4-18}) one finds
\begin{equation}
   \label{bas5-9}
\lim_{\lambda\to\infty} \tilde{Z}(N,\lambda) \; \equiv \; \tilde{Z}(s) \; = \;
\int_{-\infty}^{+\infty} dx \; \Phi(x) \; \mbox{\LARGE e}^{s x} \;
\sum_{M=1}^{\infty} \frac{(-1)^{M-1}}{(M-1)!} \; 
\biggl[\int_{x}^{+\infty} dy  \Phi(y) \biggr]^{M-1} 
\end{equation}
The summation of this series leads to the result
\begin{equation}
   \label{bas5-10}
\tilde{Z}(s) \; = \; \int_{-\infty}^{+\infty} dx \; \Phi(x) \;  
\exp\biggl(s x - \int_{x}^{+\infty} dy \Phi(y) \biggr)
\end{equation}
with the function $\Phi(x)$ is defined in Eq.\ (\ref{bas5-7}).
Note that the prefactor $N! \; \kappa^{N}$ in the expression for the full 
replica partition function, Eq.(\ref{bas4-12}), is irrelevant in the thermodynamic limit
$\lambda \to \infty$, since for fixed parameter $s = N\lambda$,
\begin{equation}
   \label{bas5-11}
N! \; \kappa^{N} \; = \; \Gamma(1+N) \; \kappa^{N} \; = \; 
\Gamma\biggl(1+\frac{s}{\lambda}\biggr) \; \kappa^{\frac{s}{\lambda}}\Bigg|_{\lambda \to \infty}
\; \to \; 1
\end{equation} 
and hence
\begin{equation}
\label{bas5-12}
\lim_{L\to\infty}
 \biggl[ Z(N,L)  \; \mbox{\LARGE e}^{\beta N L f_{0}} \biggr] \; = \; 
\tilde{Z}(s) 
\end{equation}

\section{Free Energy Distribution Function}

The distribution function of the free energy fluctuations can now be derived following
the lines of the general approach discussed in the Introduction, Eqs.(\ref{bas1-6})-(\ref{bas1-17}).
According to the definition of the replica partition function, Eq.(\ref{bas1-8}),
\begin{equation}
\label{bas6-1}
 Z(N,L)  \; = \; \int_{-\infty}^{+\infty} dF \; {\cal P}_{L}(F) \; 
\mbox{\LARGE e}^{-\beta N F} 
\end{equation}
where ${\cal P}_{L}(F)$ is the distribution function of the {\it total} free energy of the system.
Redefining the replica partition function according to Eq.(\ref{bas4-12}) and introducing
rescaled  free energy fluctuations $f$ according to the definition
\begin{equation}
\label{bas6-2}
 F \; = \; f_{0} L \; + \; \frac{1}{\beta} \lambda f
\end{equation}
(where $f_{0}$ and $\lambda$ are defined in Eqs.(\ref{bas4-15}) and (\ref{bas4-14})) instead of Eq.(\ref{bas6-1}) we get
\begin{equation}
 \label{bas6-3}
\tilde{Z}(N,L) \; = \; \int_{-\infty}^{+\infty} df \; P_{L}(f) \; 
\mbox{\LARGE e}^{-\lambda N f} 
\end{equation}
where $P_{L}(f)$ is the distribution function of the free energy fluctuations, which 
is related with ${\cal P}_{L}(F)$ via
\begin{equation}
 \label{bas6-4}
P_{L}(f) = \frac{\lambda}{\beta N! \kappa^{N}} {\cal P}_{L}(f_{0} L + \frac{\lambda}{\beta} f)
\end{equation}
Taking now the limit $L \to \infty$ in both sides of Eq.(\ref{bas6-3}),
at fixed $\lambda N \equiv s$,  we obtain
\begin{equation}
 \label{bas6-5}
\tilde{Z}(s) \; = \; \int_{-\infty}^{+\infty} df \; P_{*}(f) \;
\mbox{\LARGE e}^{-s f} 
\end{equation}
where $\tilde{Z}(s)$ is given in Eq(\ref{bas5-10}), and 
\begin{equation}
 \label{bas6-6}
P_{*}(f) \; = \; \lim_{L\to\infty} \; P_{L}(f)
\end{equation}
is the a universal thermodynamic limit distribution function of the free energy
fluctuations. This function is obtained from the relation, Eq.(\ref{bas6-5}), via inverse
Laplace transform
\begin{equation}
\label{bas6-7}
P_{*}(f) \; = \; \int_{-i\infty}^{+i\infty} \frac{ds}{2\pi i} \; 
\tilde{Z}(s) \; \mbox{\LARGE e}^{s f} 
\end{equation}
Substituting here the explicit expression for $\tilde{Z}(s)$, Eq(\ref{bas5-10}), we find
after a trivial integrations,
\begin{equation}
   \label{bas6-8}
P_{*}(f) = \Phi(-f) \;  \exp\biggl[ - \int_{-f}^{+\infty}dy \; \Phi(y) \biggr]
\end{equation}
where the function $\Phi(x)$ is defined in Eq.(\ref{bas5-7}).

\begin{figure}[htp]
\centering
   \includegraphics[width=8.0cm]{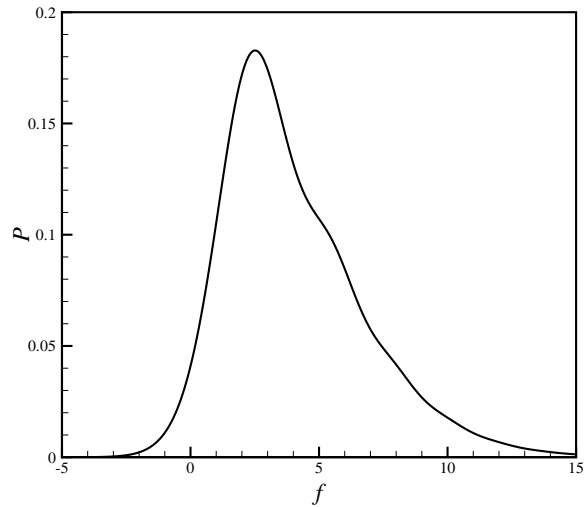}
   \caption[]{The probability distribution function $P_{*}(f)$, eq.(\ref{bas6-8})}
\label{fig1}
\end{figure}

This is the central result of the paper. The shape of the distribution function $P_{*}(f)$
is represented in Fig.1. 
One can easily check that the function $P_{*}(f)$, Eq.(\ref{bas6-8}), 
is positively defined and properly normalized. 
Indeed,  the function $\Phi(x)$, Eq.(\ref{bas5-7}),
is positive for all values of $x$ ($-\infty < x < +\infty$) and its 
asymptotics are 
\begin{eqnarray}
\label{bas6-9}
\nonumber
\Phi(x \to +\infty) \; &\sim&  \; \exp\biggl(-\frac{2}{3} x^{3/2} \biggr) \; \to \; 0 
\\
\Phi(x \to -\infty) \; &\sim&  \; |x|^{1/2} \; \to \; +\infty
\end{eqnarray}
Thus, according to Eq.(\ref{bas6-8}), 
\begin{equation}
   \label{bas6-10}
\int_{-\infty}^{+\infty}df  \; P_{*}(f) \; = \; 
-\int_{-\infty}^{+\infty}df \; \frac{d}{d f} \exp\biggl[-\int_{-f}^{+\infty}dy \; \Phi(y) \biggr] \; = \;  1
\end{equation} 
The asymptotics form of the left and  the right tails are
\begin{eqnarray}
\label{bas6-11}
P_{*}(f\to -\infty) \; &\sim&  \; \exp\biggl(-\frac{2}{3} |f|^{3/2} \biggr)  
\\
\label{bas6-12}
P_{*}(f\to +\infty) \; &\sim&  \; \exp\biggl(-c_{0} f^{3/2} \biggr)  
\end{eqnarray}
where $c_{0} \simeq 0.11$.

\section{Discussion}

Obtained result for the distribution function of the free energy fluctuations,
Eq.(\ref{bas6-8}), is rather surprising. 
At present there are exists an appreciable  list of statistical systems for which
similar distribution functions have been computed
exactly in the thermodynamic limit. These systems are: 
the polynuclear growth (PNG) model \cite{PNG_Spohn},
the longest increasing subsequences (LIS) model \cite{LIS}, 
the longest common subsequences (LCS) \cite{LCS},
the oriented digital boiling model \cite{oriented_boiling}, 
the ballistic decomposition model \cite{ballistic_decomposition}, 
and finally the zero-temperature  lattice version of the directed polymers 
with a specific (non-Gaussian) site-disorder distribution \cite{DP_johansson}.
It is remarkable that in all these systems (which are not always look similar) 
the fluctuations of the quantities which play the role
of ``energy'' are described by {\it the same} distribution function,
the so-called the Tracy-Widom (TW) distribution \cite{Tracy-Widom}.

The result obtained in the present work {\it is not} the the Tracy-Widom distribution,
although it is widely believed that the system considered here belongs to 
the same universality class as the models listed above. It is not clear, however,  which
way the term ``universality class'' should be understood.
As far as only the scaling exponents are concerned, all these systems including the present one,
indeed belong to the same universality class. On the other hand, the shapes of the 
distribution functions of the systems listed above and the result obtained for the present system 
are different.
What is really surprising is that they are different not just at the level of ``details'' but in the 
qualitative properties, such as the asymptotic behavior of the left an right tails.
The TW distribution $P_{TW}(f)$ is strongly asymmetric in its asymptotics:
\begin{eqnarray}
\label{bas7-1}
 P_{TW}(f\to -\infty) \; &\sim&  \; \exp\biggl(-\frac{2}{3} |f|^{3/2} \biggr)  
\\
\label{bas7-2}
 P_{TW}(f\to +\infty) \; &\sim&  \; \exp\biggl(-\frac{1}{12} f^{3} \biggr) 
\end{eqnarray}
hence, the {\it right tail} decays much faster than the left one. In contrast,
our distribution function is ``almost symmetric`` in its right and left tails,
cf. Eq.(\ref{bas6-11})-(\ref{bas6-12}), and moreover, due to numerical factors its {\it left tail}
decays slightly faster than the right one.

One may propose three possible explanations of the discrepancy discussed above:

(1) All the systems described by the TW distributions are essentially the zero-temperature
models, while our system by its definition is the ''high-temperature`` one. Moreover, formally,
our system has no zero-temperature limit at all. To study the limit $T \to 0$, one would need to
introduce a lattice or a ''finite width`` $\delta$-function regularization in the
model, cf. Eqs.(\ref{bas1-1})-(\ref{bas1-2}). In both cases, the Bethe anzats solution leading to
the result, Eq.(\ref{bas6-8}), would no longer be valid. On the other hand,  
from a general physical point of view the conclusion that the {\it thermodynamic limit} 
of directed polymers at $T=0$ and in the limit $T \to 0$ look different 
would be quite surprising. In particular, the solution in the
thermodynamic limit of the present system confined to a cylinder geometry
reveals no such difference \cite{Derrida1}.

(2) Unlike the present system, the disorder in all the systems described by the TW distribution 
is essentially non-Gaussian. Formally, any deviation from the Gaussian statistics of the 
disorder would again ruin our Bethe anzats solution, and at present it is not clear to what
extend this solution is ''stable`` with respect to non-Gaussian deviations.

(3) Much more likely appears the technical problem given by the third option. 
According to Eq.(\ref{bas6-5})
the distribution function $P_{*}(f)$ is defined by the replica partition function
$\tilde{Z}(s)$ at finite (both positive and negative) values of the 
parameter $s \propto L^{1/3} N$, which means that in the thermodynamic limit
$L \to \infty$ the function $P_{*}(f)$ is defined by the replica partition function
with the "number of replicas" $N \to 0$ (both positive and negative).
In the present work we have computed the replica partition function 
for arbitrary positive integer $N$ with the aim to perform an analytic continuation 
for the values $N \to 0$. Unfortunately, the analytic continuation from integers 
to arbitrary (real or complex) N is unambiguous only if the
corresponding function growth at infinity not faster than $\exp(N)$.
In our case the partition function growth as $\exp(N^3)$, 
and hence the knowledge of this function at arbitrary
integer N does not guarantee an unambiguously reconstruction 
in the region $|N| \ll 1$ \cite{thanks}. The classical example of such type of situation
is well known in the theory of the mean-field spin-glasses, such as the Sherrington and
Kirkpatrick (SK) model \cite{SK} and the Random Energy Model (REM) \cite{REM}. 
For both models one can relatively easy compute the replica partition functions 
for arbitrary positive integer number of replicas $N$, and 
in both cases the replica partition function growths as $\exp(N^2)$ at large $N$. 
Further "direct" analytic continuation of these solutions
to the region $0 < N < 1$ yields nothing else but the replica-symmetric solutions,
which at first sight look sufficiently reasonable (at least in the SK model), 
but more detailed investigation reveals that they are unphysical. 
As we know, the solution which is believed to be valid in the 
region  $0 < N < 1$ reveals the Parisi replica symmetry breaking (RSB) structure 
(one-step RSB in the case of REM), and it is derived in terms of a {\it heuristic}
procedure (directly in the interval $0 < N < 1$) and not as a proper analytic continuation 
from integer to non integer values of $N$. 
Moreover, in the case of REM  
there is a kind of the phase transition at $N=0$, which means that at negative $N$
the replica partition function should be computed separately \cite{Derrida2}.

Although up to the present moment, we have not uncovered any unphysical properties in
obtained probability function $P_{*}(f)$, the above arguments 
indicate that the present solution, Eq.\ (\ref{bas6-8}), could as well
be a kind of a distant analog of the "replica symmetric approximation", while
the derivation  of the "true" solution would require more sophisticated ideas.

\acknowledgments

$(*)$ This work was actually done in collaboration with Gianni Blatter 
(Theoretische Physik, ETH-Zurich).

We are also grateful to B.\ Derrida, M.\ Mezard, A.\ Lebedev, S.\ Nechaev, V.S.\
Dotsenko, S.\ Korshunov, and V.\ Geshkenbein for numerous fruitful discussions
of various aspects of the present work and acknowledge financial support from
the Center for Theoretical Studies at ETH Zurich and the Swiss National
Foundation.

\newpage

\begin{center}

\appendix{\Large Appendix A}

\vspace{5mm}

 {\bf \large Wave functions of quantum bosons with repulsive interactions}

\end{center}

\newcounter{A}
\setcounter{equation}{0}
\renewcommand{\theequation}{A.\arabic{equation}}

\vspace{10mm}

Explicitly the eigenstate equation (\ref{bas2-13}) reeds
\begin{equation}
 \label{A1}
\frac{1}{2\beta}\sum_{a=1}^{N}\partial_{x_a}^2 \Psi({\bf x}) \; + \; 
\frac{1}{2}\beta^2 u \sum_{a\not=b}^{N} \delta(x_a-x_b) \Psi({\bf x}) 
\; = \; - E \Psi({\bf x})
\end{equation}
Due to the symmetry of the wave function with respect to permutations of its arguments
it is sufficient to consider it in the sector 
\begin{equation}
 \label{A2}
x_{1} \; < \; x_{2} \; < \; ... \; < \; x_{N}
\end{equation}
as well as at its boundary. Inside this sector the wave function $\Psi({\bf x})$ 
satisfy the equation
\begin{equation}
 \label{A3}
\frac{1}{2\beta}\sum_{a=1}^{N}\partial_{x_a}^2 \Psi({\bf x}) 
\; = \; - E \Psi({\bf x})
\end{equation}
which describes $N$ free particles, and its generic solution is the linear combination
of $N$ plane waves characterized by $N$ momenta $\{q_{1}, q_{2}, ..., q_{N}\} \equiv {\bf q}$.
Integrating Eq.(\ref{A1}) over the variable $(x_{i+1}-x_{i})$ in a small interval
around zero, $|x_{i+1}-x_{i}| < \epsilon \to  0$, and assuming that the other
variables $\{ x_{j}\}$ (with $j \not= i, i+1$) belong to the sector, Eq.(\ref{A2}),  
one easily finds that the wave function  $\Psi({\bf x})$ must satisfy the following
boundary conditions:
\begin{equation}
\label{A4}
\bigl(\partial_{x_{i+1}} - \partial_{x_i} + \kappa \bigr) 
\Psi({\bf x})\bigg|_{x_{i+1} = x_{i} + 0} \; = \; 0
\end{equation}
where $\kappa = \beta^{3} u$.
Functions satisfying both Eq.\ (\ref{A3}) and
the boundary conditions Eq.\ (\ref{A4}) can be written in the form
\begin{equation}
\label{A5}
\Psi_{q_{1}...q_{N}}(x_{1}, ..., x_{N}) \equiv
\Psi^{(N)}_{\bf q}({\bf x}) \; = \; C 
\biggl(\prod_{a<b}^{N}\bigl[ \partial_{x_a} - \partial_{x_b} + \kappa \bigr]\biggr) \;
    \det\bigl[\widehat{\exp( i {\bf q \, x})} \bigr]
\end{equation}
where $C$ is the normalization constant to be defined later, and 
the symbol $\widehat{\exp( i {\bf q \, x})}$ denotes the $N\times N$ matrix with the 
elements $\exp( i q_a x_{b}) \; (a,b = 1,...,N)$.
First of all, it is evident that being the linear combination of the 
plane waves, the above wave function satisfy  Eq.(\ref{A3}). To demonstrate
which way this function satisfy the boundary conditions, Eq.(\ref{A4}),
let us check it, as an example, for the case $i=1$.
According to Eq.(\ref{A5}), the wave function $\Psi^{(N)}_{\bf q}({\bf x})$
can be represented in the form
\begin{equation}
\label{A6}
\Psi^{(N)}_{\bf q}({\bf x}) \; = \; 
- \bigl(\partial_{x_2} - \partial_{x_1} - \kappa \bigr) 
\tilde{\Psi}^{(N)}_{\bf q}({\bf x})
\end{equation}
where
\begin{equation}
\label{A7}
\tilde{\Psi}^{(N)}_{\bf q}({\bf x}) \; = \; C 
\biggl(\prod_{a=3}^{N}\bigl[ \partial_{x_a} - \partial_{x_1} + \kappa \bigr]
                      \bigl[ \partial_{x_a} - \partial_{x_2} + \kappa \bigr] \biggr)
\biggl(\prod_{3\leq a <b}^{N}\bigl[ \partial_{x_a} - \partial_{x_b} + \kappa \bigr]\biggr)
    \det\bigl[\widehat{\exp( i {\bf q \, x})} \bigr]
\end{equation}
One can easily see that this function is {\it antisymmetric} with respect to
the permutation of $x_{1}$ and $x_{2}$. Substituting Eq.(\ref{A6}) into Eq.(\ref{A4}) (with $i=1$)
we get
\begin{equation}
\label{A8}
-\biggl[\bigl(\partial_{x_2} - \partial_{x_1}\bigr)^{2} - \kappa^{2} \biggr] 
\tilde{\Psi}^{(N)}_{\bf q}({\bf x})\bigg|_{x_{2} = x_{1}} \; = \; 0
\end{equation}
Given the antisymmetry of the l.h.s expression with respect to
the permutation of $x_{1}$ and $x_{2}$ the above condition is indeed satisfied
at boundary $x_{1}=x_{2}$.

Since the eigenfunction $\Psi^{(N)}_{\bf q}({\bf x})$ satisfying 
Eq.(\ref{A1}) must be {\it symmetric} with respect to permutations of
its arguments, the function, Eq.(\ref{A5}), can be easily continued 
beyond the sector, Eq.(\ref{A2}), to the entire space of variables
$\{x_{1}, x_{2}, ..., x_{N}\} \in R_{N}$,
\begin{equation}
\label{A9}
\Psi^{(N)}_{\bf q}({\bf x}) \; = \; C 
\biggl(\prod_{a<b}^{N}\bigl[ -i\bigl(\partial_{x_a} - \partial_{x_b}\bigr) +i \kappa \sgn(x_{a}-x_{b})\bigr]\biggr) \;
    \det\bigl[\widehat{\exp( i {\bf q \, x})} \bigr]
\end{equation}
where, by definition, the differential operators $\partial_{x_a}$ act only on 
the exponential terms and not on the $\sgn(x)$ functions, and 
for further convenience we have redefined $i^{N(N-1)/2} C \to C$. 
Explicitly the determinant in the above equation is
\begin{equation}
\label{A10}
    \det\bigl[\widehat{\exp( i {\bf q \, x})}\bigr] \; = \; 
\sum_{P} (-1)^{[P]} \; \exp\bigl[i \sum_{a=1}^{N} q_{p_{a}} x_{a} \bigr]
\end{equation}
where the summation goes over the permutations $P$ of $N$ momenta $\{ q_{1}, q_{2}, ..., q_{N}\}$
over $N$ particles $\{ x_{1}, x_{2}, ..., x_{N}\}$, and $[P]$ denotes the parity of the permutation.
In this way the eigenfunction, Eq.(\ref{A9}), can be represented as follows
\begin{equation}
\label{A11}
\Psi^{(N)}_{\bf q}({\bf x}) \; = \; C 
\sum_{P} (-1)^{[P]} \; 
\biggl(\prod_{a<b}^{N}\bigl[ -i\bigl(\partial_{x_a} - \partial_{x_b}\bigr) +i \kappa \sgn(x_{a}-x_{b})\bigr]\biggr) \;
\exp\bigl[i \sum_{a=1}^{N} q_{p_{a}} x_{a} \bigr]
\end{equation}
 Taking the derivatives, we obtain
\begin{equation}
\label{A12}
\Psi^{(N)}_{\bf q}({\bf x}) \; = \; C 
\sum_{P} (-1)^{[P]} \; 
\biggl(\prod_{a<b}^{N}\bigl[ q_{p_a} - q_{p_b} +i \kappa \sgn(x_{a}-x_{b})\bigr]\biggr) \;
\exp\bigl[i \sum_{a=1}^{N} q_{p_{a}} x_{a} \bigr]
\end{equation}
It is evident from these representations that the eigenfunctions $\Psi^{(N)}_{\bf q}({\bf x})$ are
{\it antisymmetric} with respect to permutations of the momenta $q_{1}, ..., q_{N}$. 

Substituting the expression for the eigenfunctions, Eq.(\ref{A5}) (which is valid in
the sector, Eq.(\ref{A2})), into Eq.(\ref{A3}) for the energy spectrum we find
\begin{equation}
   \label{A20}
E \;  = \; \frac{1}{2\beta} \sum_{a=1}^{N} q_{a}^{2} 
\end{equation}

Now one can easily prove that the above eigenfunctions with different momenta are orthogonal to
each other. Let us consider two wave functions $\Psi^{(N)}_{\bf q}({\bf x})$ and 
$\Psi^{(N)}_{\bf q'}({\bf x})$ where it is assumed that
\begin{eqnarray}
 \label{A13}
q_{1} \; < \; q_{2} \; < \; ... \; < \; q_{N}\\
\nonumber
q'_{1} \; < \; q'_{2} \; < \; ... \; < \; q'_{N}
\end{eqnarray}
Using the representation, Eq.(\ref{A11}), for the overlap of these two function we get
\begin{eqnarray}
 \label{A14}
\nonumber
\overline{\Psi^{(N)^{*}}_{\bf q'}({\bf x}) \Psi^{(N)}_{\bf q}({\bf x})} &\equiv& 
\int_{-\infty}^{+\infty} d^{N}{\bf x} \; \Psi^{(N)^{*}}_{\bf q'}({\bf x}) \Psi^{(N)}_{\bf q}({\bf x}) \\
\nonumber
&=& |C|^{2} \sum_{P,P'} (-1)^{[P]+[P']}  
\int_{-\infty}^{+\infty} d^{N}{\bf x} 
\biggl\{\biggl(\prod_{a<b}^{N}
\bigl[ i\bigl(\partial_{x_a} - \partial_{x_b}\bigr) -i \kappa \sgn(x_{a}-x_{b})\bigr]\biggr) \;
\exp\bigl[-i \sum_{a=1}^{N} q'_{p'_{a}} x_{a} \bigr]\biggr\} \times \\
&\times& 
\biggl\{\biggl(\prod_{a<b}^{N}
\bigl[ -i\bigl(\partial_{x_a} - \partial_{x_b}\bigr) +i \kappa \sgn(x_{a}-x_{b})\bigr]\biggr) \;
\exp\bigl[i \sum_{a=1}^{N} q_{p_{a}} x_{a} \bigr]\biggr\}
\end{eqnarray}
Integrating by parts we obtain
\begin{eqnarray}
 \label{A15}
\overline{\Psi^{(N)^{*}}_{\bf q'}({\bf x}) \Psi^{(N)}_{\bf q}({\bf x})} &=& |C|^{2}
\sum_{P,P'} (-1)^{[P]+[P']}  
\int_{-\infty}^{+\infty} d^{N}{\bf x} \;
\exp\bigl[-i \sum_{a=1}^{N} q'_{p'_{a}} x_{a} \bigr] \times \\
\nonumber
&\times&
\biggl(\prod_{a<b}^{N}\bigl[-i\bigl(\partial_{x_a} - \partial_{x_b}\bigr) -i \kappa \sgn(x_{a}-x_{b})\bigr] \;
                      \bigl[-i\bigl(\partial_{x_a} - \partial_{x_b}\bigr) +i \kappa \sgn(x_{a}-x_{b})\bigr] \biggr)
\exp\bigl[i \sum_{a=1}^{N} q_{p_{a}} x_{a} \bigr]\biggr\}
\end{eqnarray}
or
\begin{equation}
 \label{A16}
\overline{\Psi^{(N)^{*}}_{\bf q'}({\bf x}) \Psi^{(N)}_{\bf q}({\bf x})} = |C|^{2}
\sum_{P,P'} (-1)^{[P]+[P']}  
\int_{-\infty}^{+\infty} d^{N}{\bf x} 
\exp\bigl[-i \sum_{a=1}^{N} q'_{p'_{a}} x_{a} \bigr]
\biggl(\prod_{a<b}^{N}\bigl[-(\partial_{x_a} - \partial_{x_b})^{2} + \kappa^{2} \bigr] \biggr)
\exp\bigl[i \sum_{a=1}^{N} q_{p_{a}} x_{a} \bigr]
\end{equation}
Taking the derivatives and performing the integrations
we find
\begin{eqnarray}
 \label{A17}
\nonumber
\overline{\Psi^{(N)^{*}}_{\bf q'}({\bf x}) \Psi^{(N)}_{\bf q}({\bf x})} &=& 
|C|^{2} \sum_{P,P'} (-1)^{[P]+[P']}  
\biggl(\prod_{a<b}^{N}\bigl[(q_{p_a} - q_{p_b})^{2} + \kappa^{2} \bigr] \biggr)
\int_{-\infty}^{+\infty} d^{N}{\bf x} \; 
\exp\bigl[i \sum_{a=1}^{N} (q_{p_{a}}-q'_{p'_{a}}) x_{a} \bigr] \\ 
&=& 
|C|^{2} \sum_{P,P'} (-1)^{[P]+[P']} 
\biggl(\prod_{a<b}^{N}\bigl[(q_{p_a} - q_{p_b})^{2} + \kappa^{2} \bigr] \biggr)
\biggl[\prod_{a=1}^{N} (2\pi) \delta(q_{p_a} - q'_{p'_a}) \biggr]
\end{eqnarray}
Taking into account the constraint, Eq.(\ref{A13}), one can easily note that the only the terms
which survive in the above summation over the permutations are $P = P'$, all contributing
equal value. Thus, we finally get
\begin{equation}
 \label{A18}
\overline{\Psi^{(N)^{*}}_{\bf q'}({\bf x}) \Psi^{(N)}_{\bf q}({\bf x})} = |C|^{2} \; N! \;
\biggl(\prod_{a<b}^{N}\bigl[(q_{a} - q_{b})^{2} + \kappa^{2} \bigr] \biggr)
\biggl[\prod_{a=1}^{N} (2\pi) \delta(q_{a} - q'_{a}) \biggr]
\end{equation}
With the normalization constant
\begin{equation}
   \label{A19}
|C| = \frac{1}{\sqrt{N! \prod_{a<b}^{N} 
\bigl[ (q_{a}-q_{b})^{2} +\kappa^{2} \bigr] }}
\end{equation}
we conclude that the set of the eigenfunctions, Eq.(\ref{A11}) or (\ref{A12}), are
orthonormal. The proof of completeness of this set is given in Ref. \cite{gaudin}.
It should be noted that the above wave functions present the orthonormal set of
eigenfunctions of the problem, Eq.(\ref{A1}), for any sign of the interactions
$\kappa$, e.i. both for the repulsive, $\kappa < 0$, and for the attractive, $\kappa > 0$, cases.
However, only in the case of repulsion this set is complete, while in the 
case of attractive interactions, $\kappa > 0$, in addition to the solutions, Eq.(\ref{A11}),
which describe the continuous free particles spectrum, one finds the whole family
of discrete bound eigenstates (which do not exist in the case of repulsion) 
(see Appendices B and C).


\vspace{15mm}

\begin{center}

\appendix{\Large Appendix B} 

\vspace{5mm}

{\bf \large Ground state of quantum bosons with attractive interactions}

\end{center}

\newcounter{B}
\setcounter{equation}{0}
\renewcommand{\theequation}{B.\arabic{equation}}

\vspace{10mm}

The simplest example of the bound state is the one in which all $N$ particles are bound into
one finite size ''cluster``:
\begin{equation}
   \label{B1}
   \Psi_{q}^{(1)}({\bf x}) \; = \;  C \; 
    \exp\biggl[i q \sum_{a=1}^{N} x_{a} - \frac{1}{4}\kappa \sum_{a,b=1}^{N} |x_{a}-x_{b}| \biggr]
\end{equation}
where $C$ is the normalization constant (see below) and
$q$ is the (continuous) momentum of free center of mass motion. Substituting
this function in Eq.(\ref{A1}), one can easily check that  this is indeed the eigenfunction
with the energy spectrum given by the relation
\begin{equation}
 \label{B2}
E \;  = \;  -\frac{1}{2\beta} \sum_{a=1}^{N} 
\biggl[iq - \frac{1}{2}\kappa \sum_{b=1}^{N} \sgn(x_{a}-x_{b}) \biggr]^{2}
\end{equation}
where it is assumed (by definition) that $\sgn(0) = 0$.
Since the result of the above summations does not depend on the mutual particles positions,
for simplicity we can order them according to Eq.(\ref{A2}). Then, using well known relations
\begin{eqnarray}
 \label{B3}
\sum_{b=1}^{N} \sgn(x_{a}-x_{b}) &=& -(N+1-2a)\\
\label{B4}
\sum_{a=1}^{N} a &=& \frac{1}{2} N (N+1)\\
\label{B5}
\sum_{a=1}^{N} a^{2} &=& \frac{1}{6} (N+1) (2N+1)
\end{eqnarray} 
for the energy spectrum, Eq.(\ref{B2}), we get
\begin{equation}
   \label{B6}
E \; = \; \frac{N}{2\beta} q^2 - \frac{\kappa^{2}}{24\beta}(N^{3}-N) \; \equiv \; 
E_{1}(q; N)
\end{equation}
The normalization constant $C$ is defined by the orthonormality
condition
\begin{equation}
   \label{B7}
\overline{\Psi_{q'}^{(1)^{*}}({\bf x}) \Psi_{q}^{(1)}({\bf x})} \; \equiv \;
\int_{-\infty}^{+\infty} dx_{1}...dx_{N} \; 
\Psi_{q'}^{(1)^{*}}({\bf x}) \Psi_{q}^{(1)}({\bf x}) \; = \; 
(2\pi) \delta(q-q')
\end{equation}
Substituting here Eq.(\ref{B1}) we get
\begin{eqnarray}
 \nonumber
\overline{\Psi_{q'}^{(1)^{*}}({\bf x}) \Psi_{q}^{(1)}({\bf x})} &=&
|C|^{2} \int_{-\infty}^{+\infty} dx_{1}...dx_{N} \; 
\exp\biggl[i (q-q') \sum_{a=1}^{N} x_{a} - 
\frac{1}{2}\kappa \sum_{a,b=1}^{N} |x_{a}-x_{b}| \biggr] \\
\label{B8}
&=& |C|^{2} N!
\int_{-\infty}^{+\infty} dx_{1}
\int_{x_{1}}^{+\infty} dx_{2} 
....
\int_{x_{N-1}}^{+\infty} dx_{N}
\exp\biggl[i (q-q') \sum_{a=1}^{N} x_{a} + \kappa \sum_{a=1}^{N} (N+1-2a) x_{a} \biggr]
\end{eqnarray}
where for the ordering, Eq.(\ref{A2}), we have used the relation
\begin{equation}
 \label{B9}
\frac{1}{2} \sum_{a,b=1} |x_{a}-x_{b}| \; = \; 
- \sum_{a=1}^{N} (N+1-2a) x_{a}
\end{equation}
Integrating first over $x_{N}$, then over $x_{N-1}$, and proceeding until $x_{1}$,
we find
\begin{eqnarray}
 \nonumber
\overline{\Psi_{q'}^{(1)^{*}}({\bf x}) \Psi_{q}^{(1)}({\bf x})} 
&=& |C|^{2} N! \;
\biggl(\prod_{r=1}^{N-1} \frac{1}{r[(N-r)\kappa -i(q-q')]}\biggr)
\int_{-\infty}^{+\infty} dx_{1} 
\exp\bigl[iN(q-q')x_{1}\bigr]\\
\nonumber
&=& |C|^{2} N! \; 
\biggl(\prod_{r=1}^{N-1} \frac{1}{r(N-r)\kappa}\biggr) \;
(2\pi) \delta\bigl(N(q-q')\bigr) \\
\label{B10}
&=& |C|^{2} \; \frac{N\kappa}{N!\kappa^{N}} \; (2\pi) \delta(q-q')
\end{eqnarray}
According to Eq.(\ref{B7}) this defines the normalization constant
\begin{equation}
   \label{B11}
   |C| \; = \; \sqrt{\frac{\kappa^N N!}{\kappa N}}  \; \equiv \;   C^{(1)}(q)
\end{equation}
Note that the eigenstate described by the considered wave function, Eq.(\ref{B1}), exists only
in the case of attraction, $\kappa > 0$, otherwise this function is divergent at infinity 
and consequently it is not normalizable.

It should be noted that the wave function, Eq.(\ref{B1}), can also be derived from the 
general eigenfunctions structure, Eq.(\ref{A12}), by introducing (discrete) imaginary 
parts for the momenta $q_{a}$. We assume again that the position of particles 
are ordered according to Eq.(\ref{A2}), and  define the particles' momenta
according to the rule
\begin{equation}
\label{B12}
q_{a} \; = \; q - \frac{i}{2} \kappa (N+1-2a)
\end{equation}
Substituting this into Eq.(\ref{A12}) we get
\begin{eqnarray}
  \nonumber 
  \Psi_{q}^{(1)}({x_{1} < x_{2} < ... < x_{N}}) & \propto & 
\sum_{P} (-1)^{[P]} \; 
\biggl(\prod_{a<b}^{N}\biggl[ 
  \bigl( q - \frac{i}{2} \kappa (N+1-2P_{a})\bigr) 
- \bigl( q - \frac{i}{2} \kappa (N+1-2P_{b})\bigr) -i \kappa )\biggr]\biggr) \times
\\
&&\times
\exp\biggl[ iq\sum_{a=1}^{N}x_{a} +\frac{\kappa}{2}\sum_{a=1}^{N} (N+1-2P_{a}) x_{a} \biggr]  
\\
\nonumber
\\
& \propto &
\sum_{P} (-1)^{[P]} 
\biggl(\prod_{a<b}^{N}\bigl[ P_{b}-P_{a} + 1 \bigr] \biggr) \;
\exp\biggl[ iq\sum_{a=1}^{N}x_{a} +\frac{\kappa}{2}\sum_{a=1}^{N} (N+1-2P_{a}) x_{a} \biggr] 
\label{B13}
\end{eqnarray}
Here one can easily note that due to the presence of the product 
$\prod_{a<b}^{N}[P_{b}-P_{a} + 1] $ in the summation over permutations only the 
trivial one, $P_{a} = a$, gives non-zero contribution (if we permute
any two numbers in the sequence $1, 2, ... , N$ then we can always find two numbers
$a < b$, such that $P_{b} = P_{a} - 1$). Thus
\begin{equation}
 \label{B14}
 \Psi_{q}^{(1)}({x_{1} < x_{2} < ... < x_{N}}) \; \propto \;
\exp\biggl[ iq\sum_{a=1}^{N}x_{a} +\frac{\kappa}{2}\sum_{a=1}^{N} (N+1-2a) x_{a} \biggr] 
\end{equation}
Taking into account the relation, Eq.(\ref{B9}), we recover the function, Eq.(\ref{B1}),
which is symmetric with respect to its $N$ arguments and therefore can be extended
beyond the sector, Eq.(\ref{A2}), for arbitrary particles positions.
Finally, substituting the momenta, Eq.(\ref{B12}), into the general expression for the energy 
spectrum, Eq.(\ref{A20}), we get
\begin{equation}
 \label{B15}
E \, = \; \frac{1}{2\beta} \sum_{a=1}^{N} \bigl[q - \frac{i}{2} \kappa (N+1-2a) \bigr]^{2} 
\end{equation}
Performing here simple summations (using Eqs.(\ref{B4}), (\ref{B5})) 
one recovers Eq.(\ref{B6}).


\vspace{15mm}

\begin{center}

\appendix{\Large Appendix C}

\vspace{5mm}

{\bf \large Wave functions of quantum bosons with attractive interactions}

\end{center}

\newcounter{C}
\setcounter{equation}{0}
\renewcommand{\theequation}{C.\arabic{equation}}

\vspace{10mm}

{\center{\bf 1. Eigenfunctions}}

\vspace{5mm}

The general expression for the eigenfunctions both for the case of repulsion and
for the case of attraction is given in Eqs.(\ref{bas3-1}) or (\ref{bas3-5})-(\ref{bas3-5a}).
A generic eigenfunction is characterized by $N$ momenta parameters $\{ q_{a}\} \; (a= 1,2,...N)$
which in the case of attractive interactions may have imaginary parts.
It is convenient to group these parameters into $M \; \; (1 \leq M \leq N)$ ''vector``
momenta, 
\begin{equation}
\label{C1}
q^{\alpha}_{r} \; = \; 
q_{\alpha} \; - \; \frac{i}{2} \; \kappa \; (n_{\alpha} + 1 -2 r)
\end{equation}
where $q_{\alpha} \; (\alpha = 1, 2, ..., M)$ are the continuous (real) parameters,
and the (discrete) imaginary components of each ''vector`` are labeled by an index 
$r = 1, 2, ..., n_{\alpha}$. With the given total number of particles equal to
$N$, the integers $n_\alpha$ have to satisfy the constraint
\begin{equation}
\label{C2}
\sum_{\alpha = 1}^{M} \; n_{\alpha} \; = \; N
\end{equation}
In other words, a generic eigenstate is characterized by the discrete number $M$
of complex ''vector`` momenta, by the set
of $M$ integer parameters $\{ n_{1}, n_{2}, ..., n_{M} \} \equiv {\bf n}$ (which are the numbers
of imaginary components of each ''vector``) and by the set of $M$ real continuous
momenta $\{ q_{1}, q_{2}, ..., q_{M} \}  \equiv {\bf q}$. 

To understand the structure of the determinant of the $N\times N$ matrix
$\exp( i \, q_{a} \, x_{b})$, which defines the wave functions, 
Eqs.(\ref{bas3-5})-(\ref{bas3-5a}), the $N$ momenta  
$q_{a} = q^{\alpha}_{r}$ can be ordered as follows:
\begin{equation}
 \label{C3}
\{ q_{a} \} \; = \; \{ q^{1}_{1}, \; q^{1}_{2}, \; ..., \; q^{1}_{n_{1}}; \; 
                       q^{2}_{1}, \; q^{2}_{2}, \; ..., \; q^{2}_{n_{2}}; \; ... ; \; 
                       q^{M}_{1}, \; q^{M}_{2}, \; ..., \; q^{M}_{n_{M}} \}
\end{equation}
By definition,
\begin{equation}
\label{C4}
    \det\bigl[\widehat{\exp( i {\bf q \, x})}\bigr] \; = \; 
\sum_{P} (-1)^{[P]} \; \exp\bigl[i \sum_{a=1}^{N} q_{p_{a}} x_{a} \bigr]
\end{equation}
where the summation goes over the permutations of $N$ momenta $\{ q_{a}\}$, Eq.(\ref{C3}),
over $N$ particles $\{ x_{1}, x_{2}, ..., x_{N}\}$, and $[P]$ denotes the parity of the permutation.
For a given permutation $P$ a particle number $a$ 
is attributed a momentum component $q^{\alpha(a)}_{r(a)}$.
The particles getting the momenta with the same 
 $\alpha$ (having the same real part $q_{\alpha}$) will be called  
belonging to a cluster $\Omega_{\alpha}$. For a given permutation $P$
the particles belonging 
to the same cluster are  numbered by the  "internal" index $r = 1,...,n_{\alpha}$.
Thus, according to Eq.(\ref{bas3-5a}),
\begin{equation}
 \label{C5}
\Psi_{\bf q, n}^{(M)}({\bf x}) \; = \;  C^{(M)}_{\bf q, n}
\sum_{P} (-1)^{[P]} 
    \biggl(\prod_{a<b}^{N}\bigl[ -i\bigl(\partial_{x_a} - \partial_{x_b}\bigr) +i \kappa \sgn(x_{a}-x_{b})\bigr]\biggr) \;
    \exp\biggl[ i \sum_{a=1}^{N} q^{\alpha(a)}_{r(a)} \, x_{a} \biggr]
\end{equation}
where $C^{(M)}_{\bf q, n}$ is the normalization constant to be defined later.
Substituting here Eq.(\ref{C1}) and taking derivatives we get
\begin{eqnarray}
   \label{C6}
\nonumber
  \Psi_{\bf q, n}^{(M)}({\bf x}) &=& C^{(M)}_{\bf q, n}
\sum_{P} (-1)^{[P]} 
\prod_{a<b}^{N}\biggl[ q_{\alpha(a)}- q_{\alpha(b)}
-i\kappa [\frac{n_{\alpha(a)}-n_{\alpha(b)}}{2}-r(a)+r(b)-\sgn(x_{a}-x_{b})]\biggr]  
\times\\
& &\times\exp\biggl[
i\sum_{a=1}^{N}q_{\alpha(a)} x_{a} 
 +\frac{\kappa}{2}\sum_{a=1}^{N} \bigl(n_{\alpha(a)} +1 - 2r(a)\bigr) x_{a} \biggr]
\end{eqnarray}
The pre-exponential product in the above equation contains two types of term: the pairs of 
points $(a,b)$ which belong to different clusters ($\alpha(a)\not=\alpha(b)$), and pairs of points
which belong to the same cluster ($\alpha(a) = \alpha(b)$). In the last case, the product
$\Pi_{\alpha}$ over the pairs of points which belong to a cluster $\Omega_{\alpha}$ 
reduces to 
\begin{equation}
 \label{C7}
\Pi_{\alpha} \; \propto \; 
\prod_{a<b\in\Omega_{\alpha}}\bigl[r(b) - r(a) -\sgn(x_{a}-x_{b})]\bigr] 
\end{equation}
As for the ground state wave function Eq.\ (\ref{B13})--(\ref{B14}),
one can easily note that due to the presence of this product 
in the summations over $n_{\alpha}!$ ''internal`` (inside the cluster $\Omega_{\alpha}$)
permutations $r(a)$ 
only one permutation gives non-zero contribution. 
To prove this statement, we note
that the wave function $\Psi_{\bf q, n}^{(M)}({\bf x})$ is symmetric with
respect to permutations of its $N$ arguments $\{x_{a}\}$; it is then
sufficient to consider the case where the positions of the particles are
ordered, $x_1 < x_2 < \dots < x_N$. In particular,
the particles $\{x_{a_{k}}\} \; (k = 1, 2, \dots, n_{\alpha})$
belonging to the same cluster $\Omega_{\alpha}$ are also ordered $x_{a_{1}} <
x_{a_{2}} < \dots < x_{a_{n_{\alpha}}}$.  In this case
\begin{equation}
 \label{C8}
\Pi_{\alpha} \; \propto \; 
\prod_{k<l}^{n_{\alpha}} \bigl[r(l) - r(k) + 1\bigr] 
\end{equation}
Now it is evident that the above product is non-zero only for the trivial
permutation, $r(k) = k$ (since if we permute
any two numbers in the sequence $1, 2, ... , n_{\alpha}$, we can always find two numbers
$k < l$, such that $r(l) = r(k) - 1$). In this case
\begin{equation}
 \label{C9}
\Pi_{\alpha} \; \propto \; 
\prod_{k<l}^{n_{\alpha}} \bigl[ l - k + 1 \bigr] 
\end{equation}
Including the values of all these ''internal`` products, 
Eq.(\ref{C9}), into the redefined normalization constant
$C^{(M)}_{\bf q, n}$, the wave function, Eq.(\ref{C6}) (with  $x_1 < x_2 < \dots < x_N$), reeds
\begin{eqnarray}
   \label{C10}
\nonumber
  \Psi_{\bf q, n}^{(M)}({\bf x}) &=& C^{(M)}_{\bf q, n}
{\sum_{P}}' (-1)^{[P]} 
\prod_{\substack{ a<b\\ \alpha(a)\not=\alpha(b) }}^{N}\biggl[ q_{\alpha(a)}- q_{\alpha(b)}
-i\kappa [\frac{n_{\alpha(a)}-n_{\alpha(b)}}{2}-r(a)+r(b) + 1]\biggr]  
\times\\
& &\times\exp\biggl[
i\sum_{a=1}^{N}q_{\alpha(a)} x_{a} 
 +\frac{\kappa}{2}\sum_{a=1}^{N} (n_{\alpha(a)} +1 - 2r(a)) x_{a} \biggr]
\end{eqnarray}
where the product now goes only over the pairs of particles belonging to {\it different}
clusters, and the symbol ${\sum_{P}}'$ means that the summation goes only over 
the permutations $P$ in which the "internal" indices
$r(a)$ are ordered inside each cluster.

Note that although the positions of particles belonging to the same cluster are 
ordered, the mutual positions of particles belonging to different clusters could be arbitrary,
so that geometrically the clusters are free to ''penetrate`` each other. In other words, 
the name ''cluster`` does to assume geometrically compact particles positions. 

Now using the symmetry of the wave function $\Psi_{\bf q, n}^{(M)}({\bf x})$
with respect to the permutations of its arguments the expression in Eq.(\ref{C10}) 
can be easily continued beyond the
the sector $x_1 < x_2 < ... < x_N$ for the entire coordinate space $R_{N}$.
Using the relations
\begin{equation}
 \label{C11}
\sum_{a\in\Omega_{\alpha}} (n_{\alpha} +1 - 2r(a)) x_{a} \; = \; 
\sum_{k=1}^{n_{\alpha}} (n_{\alpha} +1 - 2 k) x_{a_{k}} \; = \; 
-\frac{1}{2} \sum_{k,l=1}^{n_{\alpha}}
\big|x_{a_{k}} - x_{a_{l}}\big|
\end{equation}
and
\begin{equation}
 \label{C12}
(n_{\alpha} +1 - 2 k) \; = \; 
- \sum_{l=1}^{n_{\alpha}}
\sgn\bigl(x_{a_{k}} - x_{a_{l}}\bigr) 
\end{equation}
(where $x_{a_{1}} < x_{a_{2}} < ... < x_{a_{n_{\alpha}}}$), 
the wave function $\Psi_{\bf q, n}^{(M)}({\bf x})$, Eq.(\ref{C10}), with {\it arbitrary} particles positions reeds
\begin{eqnarray}
\nonumber
  \Psi_{\bf q, n}^{(M)}({\bf x}) &=& C^{(M)}_{\bf q, n}
{\sum_{P}}' (-1)^{[P]} 
\prod_{\substack{ a<b\\ \alpha(a)\not=\alpha(b) }}^{N}\biggl[ 
q_{\alpha(a)}- q_{\alpha(b)}
+ \frac{i\kappa}{2} \sum_{c\in\Omega_{\alpha(a)}} \sgn(x_{a}-x_{c}) 
- \frac{i\kappa}{2} \sum_{c\in\Omega_{\alpha(b)}} \sgn(x_{b}-x_{c}) 
+ i\kappa \sgn(x_{a}-x_{b}) \biggr]  
\\
& &\times\exp\biggl[
i\sum_{\alpha=1}^{M} q_{\alpha} \sum_{a\in\Omega_{\alpha}} x_{a} 
 - \frac{\kappa}{4}\sum_{\alpha=1}^{M} \sum_{a,a'\in\Omega_{\alpha}} 
\big|x_{a} - x_{a'}\big| \biggr]
\label{C13}
\end{eqnarray}
Here the summation goes only over the permutations $P$ of the momenta, Eq.(\ref{C3}),
in which "internal" indices $r(a)$ 
in the clusters are ordered according to the actual spatial ordering of the particles 
belonging to these clusters (i.e. $r(a)$ increases from the smallest $x_{a}$ in the cluster
to the largest one).
This wave function can also be re-written in the more compact form:
\begin{equation}
 \label{C14}
 \Psi_{\bf q, n}^{(M)}({\bf x}) =  C^{(M)}_{\bf q, n}
{\sum_{P}}' (-1)^{[P]} 
\prod_{\substack{ a<b\\ \alpha(a)\not=\alpha(b) }}^{N}
\biggl[ -i\bigl(\partial_{x_a} - \partial_{x_b}\bigr) 
+ i \kappa \sgn(x_{a}-x_{b})]\biggr]
\exp\biggl[
i\sum_{\alpha=1}^{M}q_{\alpha}\sum_{a\in\Omega_{\alpha}}x_{a} 
 -\frac{\kappa}{4}\sum_{\alpha=1}^{M}\sum_{a,a'\in\Omega_{\alpha}} |x_{a}-x_{a'}| \biggr]
\end{equation}

\vspace{10mm}

{\center {\bf 2. Orthogonality}}

\vspace{3mm}
We define the overlap of two wave functions characterized by the two sets of
parameters, $(M, {\bf n}, {\bf q})$ and $(M', {\bf n'}, {\bf q'})$ as
\begin{equation}
\label{C15}
Q^{(M,M')}_{{\bf n},{\bf n'}}({\bf q}, {\bf q'}) \; \equiv \; 
\int_{-\infty}^{+\infty} d^{N}{\bf x} \; 
 \Psi_{\bf q', n'}^{(M')^{*}}({\bf x})
 \Psi_{\bf q, n}^{(M)}({\bf x})  
\end{equation}
Substituting here Eq.(\ref{C14}) we get
\begin{eqnarray}
\label{C16}
Q^{(M,M')}_{{\bf n},{\bf n'}}({\bf q}, {\bf q'}) &=&
C^{(M)}_{\bf q, n} C^{(M')}_{\bf q', n'} 
{\sum_{P}}' {\sum_{P'}}' (-1)^{[P] + [P']}  \; 
\int_{-\infty}^{+\infty} d^{N}{\bf x}  
\\
\nonumber
\\
\nonumber
&& 
\biggl(
\prod_{\substack{ a<b\\ \alpha'(a)\not=\alpha'(b) }}^{N} \biggl[ 
i \bigl(\partial_{x_a} - \partial_{x_b}\bigr) - i \kappa \sgn(x_{a}-x_{b}) \biggr] \biggr)
\exp\biggl[
-i\sum_{\alpha=1}^{M'}q'_{\alpha}\sum_{a\in\Omega'_{\alpha}}^{n'_{\alpha}} x_{a} 
 -\frac{\kappa}{4}\sum_{\alpha=1}^{M'}\sum_{a,b\in\Omega'_{\alpha}}^{n'_{\alpha}} |x_{a}-x_{b}| \biggr]
\times
\\
\nonumber
\\
\nonumber
&& \times
\biggl(
\prod_{\substack{ a<b\\ \alpha(a)\not=\alpha(b) }}^{N} \biggl[ 
-i \bigl(\partial_{x_a} - \partial_{x_b}\bigr) +i \kappa \sgn(x_{a}-x_{b}) \biggr] \biggr)
\exp\biggl[
i\sum_{\alpha=1}^{M}q_{\alpha}\sum_{a\in\Omega_{\alpha}}x_{a} 
 -\frac{\kappa}{4}\sum_{\alpha=1}^{M}\sum_{a,b\in\Omega_{\alpha}} |x_{a}-x_{b}| \biggr]
\end{eqnarray}
where $\{\Omega_{\alpha}\}$ and $\{\Omega'_{\alpha}\}$ denote the clusters of the permutations $P$
and $P'$ correspondingly. Integrating by parts we obtain
\begin{eqnarray}
\nonumber
Q^{(M,M')}_{{\bf n},{\bf n'}}({\bf q}, {\bf q'}) &=&
C^{(M)}_{\bf q, n} C^{(M')}_{\bf q', n'} 
{\sum_{P}}' {\sum_{P'}}' (-1)^{[P] + [P']}  
\int_{-\infty}^{+\infty} d^{N}{\bf x}  \; 
\exp\biggl[
-i\sum_{\alpha=1}^{M'}q'_{\alpha}\sum_{a\in\Omega'_{\alpha}}^{n'_{\alpha}} x_{a} 
 -\frac{\kappa}{4}\sum_{\alpha=1}^{M'}\sum_{a,b\in\Omega'_{\alpha}}^{n'_{\alpha}} |x_{a}-x_{b}| \biggr] \times
\\
\nonumber
\\
\nonumber
&& \times
\biggl(
\prod_{\substack{ a<b\\ \alpha'(a)\not=\alpha'(b) }}^{N} \biggl[ 
-i\bigl(\partial_{x_a} - \partial_{x_b}\bigr) - i \kappa \sgn(x_{a}-x_{b}) \biggr] \biggr)
\biggl(
\prod_{\substack{ a<b\\ \alpha(a)\not=\alpha(b) }}^{N} \biggl[ 
-i\bigl(\partial_{x_a} - \partial_{x_b}\bigr) +i \kappa \sgn(x_{a}-x_{b}) \biggr] \biggr) \times
\\
\nonumber
\\
&& \times
\exp\biggl[
i\sum_{\alpha=1}^{M}q_{\alpha}\sum_{a\in\Omega_{\alpha}}x_{a} 
 -\frac{\kappa}{4}\sum_{\alpha=1}^{M}\sum_{a,b\in\Omega_{\alpha}} |x_{a}-x_{b}| \biggr]
\label{C17}
\end{eqnarray}

First, let us consider the case when the integer parameters of the two functions
coincide, $M = M'$ and ${\bf n} = {\bf n'}$, and for the moment let us suppose that 
all these integer parameters $\{n_{\alpha}\}$ are {\it different},
$1 \leq n_{1} < n_{2} < ... < n_{M}$. Then, in the summations over the permutations
in Eq.(\ref{C17}), we find two types of terms:

(A) the "diagonal" ones in which the two
permutations coincide, $P = P'$ ;

(B) the "off-diagonal" ones in which
the two permutations are different, $P \not= P'$.

The contribution of the ''diagonal`` ones reeds
\begin{eqnarray}
\nonumber
Q^{(M,M)^{(A)}}_{{\bf n},{\bf n}}({\bf q}, {\bf q'}) &=&
C^{(M)}_{\bf q, n} C^{(M)}_{\bf q', n} \;
{\sum_{P}}'
\int_{-\infty}^{+\infty} d^{N}{\bf x}  \; 
\exp\biggl[
-i\sum_{\alpha=1}^{M} q'_{\alpha}\sum_{a\in\Omega_{\alpha}}^{n_{\alpha}} x_{a} 
 -\frac{\kappa}{4}\sum_{\alpha=1}^{M}\sum_{a,b\in\Omega_{\alpha}}^{n_{\alpha}} |x_{a}-x_{b}| \biggr] \times
\\
\nonumber
\\
&& \times
\biggl(
\prod_{\substack{ a<b\\ \alpha(a)\not=\alpha(b) }}^{N} \biggl[ 
-\bigl(\partial_{x_a} - \partial_{x_b}\bigr)^{2} + \kappa^{2} \biggr] \biggr) \;
\exp\biggl[
i\sum_{\alpha=1}^{M}q_{\alpha}\sum_{a\in\Omega_{\alpha}}x_{a} 
 -\frac{\kappa}{4}\sum_{\alpha=1}^{M}\sum_{a,b\in\Omega_{\alpha}} |x_{a}-x_{b}| \biggr]
\label{C18}
\end{eqnarray}
It is evident that all permutations $\alpha(a)$ in the above equation give the same contribution
and therefore it is sufficient to consider only the contribution of the ''trivial``
permutation which is represented by line in Eq.(\ref{C3}). The cluster  ordering
given by this permutation we denote by $\alpha_{0}(a)$. For this particular configuration
of clusters we can redefine the particles numbering, so that instead of a ''plane``
index $a = 1, 2, ..., N$ the particles would be counted by two indices $(\alpha, r)$:
$\{x_{a}\} \to \{x_{r}^{\alpha}\} \; (\alpha = 1, ..., M) \; (r = 1, ..., n_{\alpha})$
indicating to which cluster $\alpha$ a given particle belongs and what is
its ''internal`` cluster number $r$. Due to the symmetry of the integrated expression
in Eq.(\ref{C18}) with respect to the permutations of the particles inside the clusters,
we can introduce the ''internal`` particles ordering for every cluster: 
$x_{1}^{\alpha} < x_{2}^{\alpha} < ... < x_{n_{\alpha}}^{\alpha}$. In this way,
using the relation, Eq.(\ref{C11}), we get
\begin{eqnarray}
\nonumber
Q^{(M,M)^{(A)}}_{{\bf n},{\bf n}}({\bf q}, {\bf q'}) &=&
C^{(M)}_{\bf q, n} C^{(M)}_{\bf q', n} \; \frac{N!}{n_{1}! n_{2}! ... n_{M}!}
\biggl[\prod_{\alpha=1}^{M}
\biggl(
n_{\alpha}!
\int_{-\infty}^{+\infty} dx_{1}^{\alpha}
\int_{x_{1}^{\alpha}}^{+\infty} dx_{2}^{\alpha}
....
\int_{x_{n_{\alpha}-1}^{\alpha}}^{+\infty} dx_{n_{\alpha}}^{\alpha}
\biggr) \biggr] \times
\\
\nonumber
\\
\nonumber
&& \times
\exp\biggl[
-i\sum_{\alpha=1}^{M} q'_{\alpha}\sum_{r=1}^{n_{\alpha}} x_{r}^{\alpha} 
+\frac{\kappa}{2}\sum_{\alpha=1}^{M}\sum_{r=1}^{n_{\alpha}} (n_{\alpha} +1 - 2r)x_{r}^{\alpha} \biggr] \times
\\
\nonumber
\\
\nonumber
&& \times
\biggl(
\prod_{\alpha < \beta}^{M} \prod_{r=1}^{n_{\alpha}} \prod_{r'=1}^{n_{\beta}}
\biggl[ 
-\bigl(\partial_{x_{r}^{\alpha}} - \partial_{x_{r'}^{\beta}}\bigr)^{2} + \kappa^{2} \biggr] \biggr) \times
\\
\nonumber
\\
&& \times
\exp\biggl[
i\sum_{\alpha=1}^{M} q_{\alpha}\sum_{r=1}^{n_{\alpha}} x_{r}^{\alpha} 
+\frac{\kappa}{2}\sum_{\alpha=1}^{M}\sum_{r=1}^{n_{\alpha}} (n_{\alpha} +1 - 2r)x_{r}^{\alpha}\biggr]
\label{C19}
\end{eqnarray}
where the factor $N!/n_{1}!...n_{M}!$ is the total number of permutations of $M$ clusters
over $N$ particles. Taking the derivatives and reorganizing the terms we obtain
\begin{eqnarray}
\nonumber
Q^{(M,M)^{(A)}}_{{\bf n},{\bf n}}({\bf q}, {\bf q'}) &=&
C^{(M)}_{\bf q, n} C^{(M)}_{\bf q', n} \; N! \; 
\biggl(
\prod_{\alpha < \beta}^{M} \prod_{r=1}^{n_{\alpha}} \prod_{r'=1}^{n_{\beta}}
\biggl[ 
\bigl[q_{\alpha} - q_{\beta} -\frac{i\kappa}{2} (n_{\alpha} - n_{\beta} - 2r +2r')\bigr]^{2} + \kappa^{2} 
\biggr] \biggr) \times
\\
\nonumber
\\
&& \times
\prod_{\alpha=1}^{M} \Biggr\{
\int_{-\infty}^{+\infty} dx_{1}^{\alpha}
\int_{x_{1}^{\alpha}}^{+\infty} dx_{2}^{\alpha}
....
\int_{x_{n_{\alpha}-1}^{\alpha}}^{+\infty} dx_{n_{\alpha}}^{\alpha} \; 
\mbox{\Large e}^{ \;
i(q_{\alpha}-q'_{\alpha}) \sum_{r=1}^{n_{\alpha}} x_{r}^{\alpha} 
+\kappa\sum_{r=1}^{n_{\alpha}} (n_{\alpha} +1 - 2r)x_{r}^{\alpha}}
\Biggr\}
\label{C20}
\end{eqnarray}
Simple integrations over $x_{r}^{\alpha}$ yields (cf. Eqs.(\ref{B8})-(\ref{B11}))
\begin{eqnarray}
\nonumber
Q^{(M,M)^{(A)}}_{{\bf n},{\bf n}}({\bf q}, {\bf q'}) &=&
\bigl(C^{(M)}_{\bf q, n}\bigr)^{2}  \; N! \; 
\biggl(
\prod_{\alpha < \beta}^{M} \prod_{r=1}^{n_{\alpha}} \prod_{r'=1}^{n_{\beta}}
\biggl[ 
\bigl[q_{\alpha} - q_{\beta} -\frac{i\kappa}{2} (n_{\alpha} - n_{\beta} - 2r +2r')\bigr]^{2} + \kappa^{2} 
\biggr] \biggr) \times
\\
\nonumber
\\
&& \times
\prod_{\alpha=1}^{M} \Biggr[
\frac{n_{\alpha}\kappa}{(n_{\alpha}!)^{2} \kappa^{n_{\alpha}}} \; 
(2\pi) \delta(q_{\alpha} - q'_{\alpha}) 
\Biggr]
\label{C21}
\end{eqnarray}

Now let us prove that the  ''off-diagonal`` terms of Eq.(\ref{C17}),
in which the permutations $P$ and $P'$ are different, give no contribution.
Here we can also chose one of the permutations, say the permutation $P$, to be the ''trivial`` one 
represented by line in Eq.(\ref{C3}) with the cluster  ordering
denoted by $\alpha_{0}(a)$. Given the symmetry of the wave functions 
it will be sufficient to consider the contribution of the sector
$x_{1} < x_{2} < ... < x_{n_{N}}$.
According to Eq.(\ref{C17}), we get
\begin{eqnarray}
\nonumber
Q^{(M,M)^{(B)}}_{{\bf n},{\bf n}}({\bf q}, {\bf q'}) &\propto&
 {\sum_{P'}}' (-1)^{[P']}  
\int_{x_{1} < ... < x_{n_{N}}} d^{N}{\bf x}  \; 
\exp\biggl[
-i\sum_{\alpha=1}^{M}q'_{\alpha}\sum_{a\in\Omega'_{\alpha}}^{n_{\alpha}} x_{a} 
 -\frac{\kappa}{4}\sum_{\alpha=1}^{M}\sum_{a,b\in\Omega'_{\alpha}}^{n_{\alpha}} |x_{a}-x_{b}| \biggr] \times
\\
\nonumber
\\
\nonumber
&& \times
\biggl(
\prod_{\substack{ a<b\\ \alpha_{0}(a)\not=\alpha_{0}(b) }}^{N} \biggl[ 
-i\bigl(\partial_{x_a} - \partial_{x_b}\bigr) +i \kappa \sgn(x_{a}-x_{b}) \biggr] \biggr)
\biggl(
\prod_{\substack{ a<b\\ \alpha'(a)\not=\alpha'(b) }}^{N} \biggl[ 
-i\bigl(\partial_{x_a} - \partial_{x_b}\bigr) -i \kappa \sgn(x_{a}-x_{b}) \biggr] \biggr)
\\
\nonumber
\\
&& \times
\exp\biggl[
i\sum_{\alpha=1}^{M}q_{\alpha}\sum_{a\in\Omega^{o}_{\alpha}}x_{a} 
 +\frac{\kappa}{2}\sum_{\alpha=1}^{M}\sum_{a\in\Omega^{o}_{\alpha}} (n_{\alpha}+1-2r(a))x_{a} \biggr]
\label{C22}
\end{eqnarray}
Here the symbols $\{\Omega^{o}_{\alpha}\}$  denote the clusters 
of the trivial permutation $\alpha_{0}(a)$.
Since $P' \not= P$,
some of the clusters $\Omega'_{\alpha}$ must be different from $\Omega^{o}_{\alpha}$.
As an illustration, let us consider a particular case of $N=10$, with three clusters
$n_{1}=5$ (denoted by the symbol ''$\bigcirc$``) , $n_{2}=2$ (denoted by the symbol ''$\times$``)
and $n_{3}=3$ (denoted by the symbol '$\triangle$``):

\vspace{5mm}

\begin{center}

\begin{tabular}{|c||c|c|c|c|c|c|c|c|c|c|}
 \hline
particle number $a$ & 1 & 2 & 3 & 4 & 5 & 6 & 7 & 8 & 9 & 10 \\
\hline
permutation $\alpha_{0}(a)$ &$\bigcirc$&$\bigcirc$&$\bigcirc$&$\bigcirc$&$\bigcirc$&$\times$&$\times$&$\triangle$&$\triangle$&$\triangle$ \\
\hline
permutation $\alpha'(a)$ &$\bigcirc$&$\bigcirc$&$\bigcirc$&$\triangle$&$\bigcirc$&$\times$&$\times$&$\bigcirc$&$\triangle$&$\triangle$ \\
\hline
\end{tabular}

\end{center}

\vspace{5mm}
Here in the permutation $\alpha'(a)$ the particle $a=4$ belong to the cluster $\alpha=3$ 
(and not to the cluster $\alpha=1$ as in the permutation $\alpha_{0}(a)$), and the 
particle $a=8$ belong to the cluster $\alpha=1$ (and not to the cluster $\alpha=3$
as in the permutation $\alpha_{0}(a)$). Now let us look carefully at the structure of the 
products in Eq.(\ref{C22}). Unlike the first product, which contains no ''internal``
products among particles belonging to the cluster $\Omega^{o}_{1}$, the second product
{\it does}. Besides, the signs of the differential operators
$\bigl(\partial_{x_a} - \partial_{x_b}\bigr)$ in the second product 
is {\it opposite} to the ''normal`` ones in the first product (cf. Eqs.(\ref{C7})-(\ref{C9})).
It is these two factors (the presence of the ''internal`` products and the ''wrong`` signs
of the differential operators) which makes the ''off-diagonal`` contributions, Eq.(\ref{C22}),
to be zero. Indeed, in the above example, the second product contains the term
\begin{equation}
 \label{C23}
\Pi'_{4,5} \; \equiv \; 
\biggl[-i\bigl(\partial_{x_4} - \partial_{x_5}\bigr) + i \kappa  \biggr] 
\exp\biggl[
i\sum_{\alpha=1}^{3} q_{\alpha}\sum_{a\in\Omega^{o}_{\alpha}} x_{a} 
 +\frac{\kappa}{2}\sum_{\alpha=1}^{3}\sum_{a\in\Omega^{o}_{\alpha}} (n_{\alpha}+1-2r(a))x_{a} \biggr]
\end{equation}
(we remind that the particles in the clusters $\Omega^{o}_{\alpha}$ are ordered,
and in particular  $x_{4} < x_{5}$).
Taking the derivatives, we get
\begin{eqnarray}
 \nonumber
\Pi'_{4,5}  &\propto&  
\biggl[-\biggl(iq_{1} + \frac{\kappa}{2}\bigl(n_{1}+1-2r(4)\bigr) 
             - iq_{1} - \frac{\kappa}{2}\bigl(n_{1}+1-2r(5)\bigr) \biggr) + \kappa \Biggr] \\
&\propto&
\bigl[ r(4)-r(5) + 1 \bigr] \; = \; 0
\label{C24}
\end{eqnarray}
since in the first cluster $r(a)=a$. 

One can easily understand that the above example reflect 
the general situation. Since all the cluster sizes $n_{\alpha}$ are supposed to be different,
whatever the permutation $\alpha'(a)$ is, we can always find a cluster
$\Omega^{o}_{\alpha}$ 
such that some of its particles belong to the same cluster number $\alpha$ in the permutation
$\alpha'(a)$ while the others do not. Then one has to consider the contribution of
the product of two {\it neighboring number} points
\begin{equation}
 \label{C25}
\Pi'_{k, k+1} \; = \; 
\biggl[-i\bigl(\partial_{x_k} - \partial_{x_{k+1}}\bigr) + i \kappa  \biggr] 
\exp\biggl[
i\sum_{\alpha=1}^{M} q_{\alpha}\sum_{a\in\Omega^{o}_{\alpha}} x_{a} 
 +\frac{\kappa}{2}\sum_{\alpha=1}^{M}\sum_{a\in\Omega^{o}_{\alpha}} (n_{\alpha}+1-2r(a))x_{a} \biggr]
\end{equation}
where in the permutation $\alpha'(a)$ the particle $k$ belong to the cluster number $\alpha$ 
and the particle $(k+1)$ belong to some
other cluster. Taking the derivatives
one gets
\begin{equation}
 \label{C26}
\Pi'_{k, k+1} \; \propto \; 
\bigl[r(k)-r(k+1) + 1\bigr] \; = \; 0
\end{equation}
as $r(a)$ is the ''internal`` particle number in the cluster $\Omega^{o}_{\alpha}$,
where $r(k+1) = r(k) + 1$ (cf. Eqs.(\ref{C7})-(\ref{C9})).

Thus, the only non-zero contribution to the overlap, Eq.(\ref{C15}), of two wave function 
$\Psi_{\bf q', n}^{(M)}({\bf x})$ and $\Psi_{\bf q, n}^{(M)}({\bf x})$ (having the same
number of clusters $M$ and characterized by
the same set of the integer parameters  $1 \leq n_{1} < n_{2} < ... < n_{M}$) comes from the ''diagonal`` terms, Eq.(\ref{C21}):
\begin{eqnarray}
\nonumber
Q^{(M,M)}_{{\bf n},{\bf n}}({\bf q}, {\bf q'}) &=&
\bigl(C^{(M)}_{\bf q, n}\bigr)^{2}  \; N! \; 
\prod_{\alpha=1}^{M} \Biggr[
\frac{n_{\alpha}\kappa}{(n_{\alpha}!)^{2} \kappa^{n_{\alpha}}} 
\Biggr] \times
\\
\nonumber
\\
&& \times
\biggl(
\prod_{\alpha < \beta}^{M} \prod_{r=1}^{n_{\alpha}} \prod_{r'=1}^{n_{\beta}}
\biggl[ 
\bigl[q_{\alpha} - q_{\beta} -\frac{i\kappa}{2} (n_{\alpha} - n_{\beta} - 2r +2r')\bigr]^{2} + \kappa^{2} 
\biggr] \biggr) \;
\prod_{\alpha=1}^{M} \Biggr[
(2\pi) \delta(q_{\alpha} - q'_{\alpha}) 
\Biggr]
\label{C27}
\end{eqnarray}

The situation when there are clusters which have the same numbers of particles $n_{\alpha}$ 
is somewhat more complicated. Let us consider the overlap between two wave function
$\Psi_{\bf q', n}^{(M)}({\bf x})$ and $\Psi_{\bf q, n}^{(M)}({\bf x})$ (which, as before
have the same $M$ and ${\bf n}$) such that in the set of $M$ integers
$n_{1}, n_{2}, ... , n_{M}$ there are two $n_{\alpha}$'s which are equal, say
$n_{\alpha_{1}} = n_{\alpha_{2}}$ (where $\alpha_{1} \not= \alpha_{2}$).
In the eigenstate $({\bf q', n})$ these two clusters have 
the center of mass momenta $q'_{\alpha_{1}}$ and $q'_{\alpha_{2}}$, and in the 
the eigenstate $({\bf q, n})$ they have the momenta $q_{\alpha_{1}}$ and $q_{\alpha_{2}}$
correspondingly.
According to the above discussion, the non-zero contributions in the summation
over the cluster permutations $\alpha(a)$ and $\alpha'(a)$ in Eq.(\ref{C17})
appears only if the clusters $\{\Omega_{\alpha}\}$ of the permutation $\alpha(a)$
totally coincide with the clusters $\{\Omega'_{\alpha}\}$ of the permutation $\alpha'(a)$.
In the case when all $n_{\alpha}$ are different this is possible only if 
the permutation $\alpha(a)$ coincides with the permutation $\alpha'(a)$. In contrast to that,
in the case when we have $n_{\alpha_{1}} = n_{\alpha_{2}}$, there are {\it two}
non-zero options. The first one, as before, is given by the ''diagonal`` terms
with $\alpha(a) = \alpha'(a)$ (so that the clusters $\{\Omega_{\alpha}\}$ and
$\{\Omega'_{\alpha}\}$ are just the same), and this contribution
is proportional to $\delta(q_{\alpha_{1}} - q'_{\alpha_{1}}) \, \delta(q_{\alpha_{2}} - q'_{\alpha_{2}})$.
The second  (''off-diagonal``) contribution is given by such permutation $\alpha'(a)$
in which the cluster $\Omega'_{\alpha_{1}}$ (of the permutation $\alpha'(a)$) coincide
with the cluster $\Omega_{\alpha_{2}}$ (of the permutation $\alpha(a)$) and 
the cluster $\Omega'_{\alpha_{2}}$ (of the permutation $\alpha'(a)$) coincide
with the cluster $\Omega_{\alpha_{1}}$ (of the permutation $\alpha(a)$) while the 
rest of the clusters of these two permutations are the same, 
$\Omega'_{\alpha} = \Omega_{\alpha} \; (\alpha \not= \alpha_{1}, \alpha_{2})$.
Correspondingly, this last contribution is proportional to 
$\delta(q_{\alpha_{1}} - q'_{\alpha_{2}}) \, \delta(q_{\alpha_{2}} - q'_{\alpha_{1}})\; (-1)^{n_{\alpha_{1}}}$.
In fact this situation with two equivalent contributions is the consequence of the symmetry
of the wave function $\Psi_{\bf q', n}^{(M)}({\bf x})$: 
the permutation of two momenta $q_{\alpha_{1}}$ and $q_{\alpha_{2}}$ belonging to 
the clusters which have the same numbers of particles, 
$n_{\alpha_{1}}=n_{\alpha_{2}}$ produces just the factor $(-1)^{n_{\alpha_{1}}}$
(see discussion below Eq.(\ref{bas3-19})). Therefore considering the clusters with equal numbers of particles
as equivalent and restricting analysis to the sectors 
$q_{\alpha_{1}} < q_{\alpha_{2}}; \; q'_{\alpha_{1}} < q'_{\alpha_{2}}$
we find that the second contribution,  
$\delta(q_{\alpha_{1}} - q'_{\alpha_{2}}) \, \delta(q_{\alpha_{2}} - q'_{\alpha_{1}})$
is identically equal to zero, thus returning to the above result Eq.(\ref{C27}).

In a generic case the $M$-component vector ${\bf n}$ can be represented in the form
\begin{equation}
 \label{C28}
{\bf n} \; = \;   \{\underbrace{m_{1}, ..., m_{1}}_{s_{1}}, 
                  \underbrace{m_{2}, ..., m_{2}}_{s_{2}}, ... ..., 
                  \underbrace{m_{k}, ..., m_{k}}_{s_{k}} \}
\end{equation}
where  $s_{1}+s_{2}+ ... +s_{k} = M$ and $k$ integers $\{ m_{i}\}$ ($ 1\leq k \leq M$)
are all supposed to be {\it different}: 
\begin{equation}
 \label{C29}
1 \leq m_{1} < m_{2} < ... < m_{k}
\end{equation}
Due to the symmetry  with respect to the momenta permutations inside the subsets 
of equal $n$'s it is sufficient to consider the wave functions in the sectors
\begin{eqnarray}
\label{C30}
&&
q_{1} < q_{2} < ... < q_{s_{1}} \; ;\\
\nonumber
&&
q_{s_{1}+1} < q_{s_{1}+2} < ... < q_{s_{1}+s_{2}} \; ; \\
\nonumber
&&
................. \\
\nonumber
&&
q_{s_{1}+...+s_{k-1}+1} < q_{s_{1}+...+s_{k-1}+2} < ... < q_{s_{1}+...+s_{k-1}+s_{k}}
\end{eqnarray}
In this representation we again recover the above result Eq.(\ref{C27})

Finally, let us consider the overlap of two eigenstates described by two {\it different}
sets of integer parameters,  ${\bf n'} \not= {\bf n}$. In fact this situation is quite simple
because if the clusters of the two states are different from each other, it means
that in the summation over the pairs of permutations $P$ and $P'$ in Eq.(\ref{C17})
there exist no two permutations for which these two sets of clusters $\{\Omega_{\alpha}\}$ and 
$\{\Omega'_{\alpha}\}$ would coincide. Which, according to the above analysis, means that 
this expression is equal to zero. Note that the condition $M' \not= M$ automatically implies 
that ${\bf n'} \not= {\bf n}$. 

Thus we have proved that
\begin{eqnarray}
\nonumber
Q^{(M,M')}_{{\bf n},{\bf n'}}({\bf q}, {\bf q'}) &=& 
\bigl(C^{(M)}_{\bf q, n}\bigr)^{2}  \;
{\boldsymbol \delta}(M,M') \; \biggl(\prod_{\alpha=1}^{M} 
{\boldsymbol \delta}(n_{\alpha},n'_{\alpha}) \biggr)
\biggl(\prod_{\alpha=1}^{M} (2\pi) \delta(q_{\alpha}-q'_{\alpha}) \biggr)
\times
\\
\nonumber
\\
&& \times
 N! \; 
\prod_{\alpha=1}^{M} \Biggr[
\frac{n_{\alpha}\kappa}{(n_{\alpha}!)^{2} \kappa^{n_{\alpha}}} 
\Biggr]
\biggl(
\prod_{\alpha < \beta}^{M} \prod_{r=1}^{n_{\alpha}} \prod_{r'=1}^{n_{\beta}}
\biggl[ 
\bigl[q_{\alpha} - q_{\beta} -\frac{i\kappa}{2} (n_{\alpha} - n_{\beta} - 2r +2r')\bigr]^{2} + \kappa^{2} 
\biggr] \biggr) 
\label{C31}
\end{eqnarray}
where the integer parameters $\{n_{\alpha}\}$ and $\{n'_{\alpha}\}$ are assumed to have the 
generic structure represented in Eqs.(\ref{C28})-(\ref{C29}), and the  momenta 
$\{q_{\alpha}\}$ and $\{q'_{\alpha}\}$ of the clusters with equal numbers of particles 
are restricted in the sectors, Eq.(\ref{C30}). 
Finally, according to Eq.(\ref{C31}), the orthonormality condition defines the normalization
constant $C^{(M)}_{\bf q, n}$ as it is given in Eq.(\ref{bas3-27}).


\vspace{15mm}

\begin{center}
\appendix{\Large Appendix D} 
\end{center}

\newcounter{D}
\setcounter{equation}{0}
\renewcommand{\theequation}{D.\arabic{equation}}

\vspace{10mm}

In this Appendix we simplify the last term in the r.h.s of  Eq.(\ref{bas4-11}),
and prove that
\begin{equation}
\label{D1}
{\boldsymbol \Pi} \; \equiv \; 
\prod_{r=1}^{n_{\alpha}} \prod_{r'=1}^{n_{\beta}}
\frac{\big|q_{\alpha}-q_{\beta} -\frac{i\kappa}{2}(n_{\alpha}-n_{\beta}-2r+2r')\big|^{2}}{
     \bigl[q_{\alpha}-q_{\beta} -\frac{i\kappa}{2}(n_{\alpha}-n_{\beta}-2r+2r')\bigr]^{2} + \kappa^{2}} \; = \; 
\frac{\big|q_{\alpha}-q_{\beta} -\frac{i\kappa}{2}(n_{\alpha}-n_{\beta})\big|^{2}}{
      \big|q_{\alpha}-q_{\beta} -\frac{i\kappa}{2}(n_{\alpha}+n_{\beta})\big|^{2}}
\end{equation}
We rewrite ${\boldsymbol \Pi}$ as
\begin{equation}
\label{D2}
{\boldsymbol \Pi} \; = \; 
\frac{
\prod_{r=1}^{n_{\alpha}} \prod_{r'=1}^{n_{\beta}}
\big|(q_{\alpha} -\frac{i\kappa}{2} n_{\alpha}) - (q_{\beta} -\frac{i\kappa}{2} n_{\beta})
    +i\kappa (r - r') \big|^{2}}{
\prod_{r=1}^{n_{\alpha}} \prod_{r'=1}^{n_{\beta}}
\bigl[(q_{\alpha} -\frac{i\kappa}{2} n_{\alpha} +i\kappa r) -
      (q_{\beta} -\frac{i\kappa}{2} n_{\beta} +i\kappa r') -i\kappa\bigr]
\bigl[(q_{\alpha} -\frac{i\kappa}{2} n_{\alpha} +i\kappa r) -
      (q_{\beta} -\frac{i\kappa}{2} n_{\beta} +i\kappa r') +i\kappa\bigr]}
\end{equation}
Redefining the indices $r$ and $r'$ of the product in the {\it left} brackets
 $\bigl[ ... \bigr]$, of the  denominator 
\begin{eqnarray}
\label{D3}
r &\rightarrow& n_{\alpha} + 1 - r\\
\nonumber
r' &\rightarrow& n_{\beta} + 1 - r'
\end{eqnarray}
we find
\begin{equation}
\label{D4}
{\boldsymbol \Pi} \; = \; 
\frac{
\prod_{r=1}^{n_{\alpha}} \prod_{r'=1}^{n_{\beta}}
\big|(q_{\alpha} -\frac{i\kappa}{2} n_{\alpha}) - (q_{\beta} -\frac{i\kappa}{2} n_{\beta})
    +i\kappa (r - r') \big|^{2}}{
\prod_{r=1}^{n_{\alpha}} \prod_{r'=1}^{n_{\beta}}
\bigl[(q_{\alpha} +\frac{i\kappa}{2} n_{\alpha}) -
      (q_{\beta} +\frac{i\kappa}{2} n_{\beta}) - i\kappa(r - r' + 1)\bigr]
\bigl[(q_{\alpha} -\frac{i\kappa}{2} n_{\alpha}) -
      (q_{\beta} -\frac{i\kappa}{2} n_{\beta}) + i\kappa(r - r' + 1)\bigr]}
\end{equation}
or
\begin{equation}
\label{D5}
{\boldsymbol \Pi} \; = \; 
\frac{
\prod_{r=1}^{n_{\alpha}} \prod_{r'=1}^{n_{\beta}}
\big|(q_{\alpha} -\frac{i\kappa}{2} n_{\alpha}) - (q_{\beta} -\frac{i\kappa}{2} n_{\beta})
    +i\kappa (r - r') \big|^{2}}{
\prod_{r=1}^{n_{\alpha}} \prod_{r'=1}^{n_{\beta}}
\big|(q_{\alpha} -\frac{i\kappa}{2} n_{\alpha}) -
      (q_{\beta} -\frac{i\kappa}{2} n_{\beta}) + i\kappa(r - r' + 1)\big|^{2}}
\end{equation}
Now, shifting the product over $r$ in the denominator by $1$ we obtain
\begin{equation}
\label{D6}
{\boldsymbol \Pi} \; = \; 
\frac{
\prod_{r=1}^{n_{\alpha}} \prod_{r'=1}^{n_{\beta}}
\big|(q_{\alpha} -\frac{i\kappa}{2} n_{\alpha}) - (q_{\beta} -\frac{i\kappa}{2} n_{\beta})
    +i\kappa (r - r') \big|^{2}}{
\prod_{r=2}^{n_{\alpha}+1} \prod_{r'=1}^{n_{\beta}}
\big|(q_{\alpha} -\frac{i\kappa}{2} n_{\alpha}) -
      (q_{\beta} -\frac{i\kappa}{2} n_{\beta}) + i\kappa(r - r')\big|^{2}} \; = \; 
\frac{
\prod_{r'=1}^{n_{\beta}}
\big|(q_{\alpha} -\frac{i\kappa}{2} n_{\alpha}) - (q_{\beta} -\frac{i\kappa}{2} n_{\beta})
    +i\kappa (1 - r') \big|^{2}}{
\prod_{r'=1}^{n_{\beta}}
\big|(q_{\alpha} -\frac{i\kappa}{2} n_{\alpha}) -
      (q_{\beta} -\frac{i\kappa}{2} n_{\beta}) + i\kappa(n_{\alpha} +1 - r')\big|^{2}}
\end{equation}
Redefining $r'$ in the product in the denominator,
\begin{equation}
\label{D7}
r' \; \rightarrow \; n_{\beta} + 1 - r'
\end{equation}
we obtain
\begin{equation}
\label{D8}
{\boldsymbol \Pi} \; = \; 
\frac{
\prod_{r'=1}^{n_{\beta}}
\big|(q_{\alpha} -\frac{i\kappa}{2} n_{\alpha}) - (q_{\beta} -\frac{i\kappa}{2} n_{\beta})
    - i\kappa (r' - 1) \big|^{2}}{
\prod_{r'=1}^{n_{\beta}}
\big|(q_{\alpha} +\frac{i\kappa}{2} n_{\alpha}) -
      (q_{\beta} +\frac{i\kappa}{2} n_{\beta}) + i\kappa  r'\big|^{2}} \; = \; 
\frac{
\prod_{r'=1}^{n_{\beta}}
\big|(q_{\alpha} -\frac{i\kappa}{2} n_{\alpha}) - (q_{\beta} -\frac{i\kappa}{2} n_{\beta})
    - i\kappa (r' - 1) \big|^{2}}{
\prod_{r'=1}^{n_{\beta}}
\big|(q_{\alpha} -\frac{i\kappa}{2} n_{\alpha}) - (q_{\beta} -\frac{i\kappa}{2} n_{\beta}) 
    - i\kappa  r'\big|^{2}}
\end{equation}
Finally, shifting the product in the numerator by $(-1)$, we get
\begin{equation}
\label{D9}
{\boldsymbol \Pi} \; = \; 
\frac{
\prod_{r'=0}^{n_{\beta}-1}
\big|(q_{\alpha} -\frac{i\kappa}{2} n_{\alpha}) - (q_{\beta} -\frac{i\kappa}{2} n_{\beta})
    - i\kappa r' \big|^{2}}{
\prod_{r'=1}^{n_{\beta}}
\big|(q_{\alpha} -\frac{i\kappa}{2} n_{\alpha}) - (q_{\beta} -\frac{i\kappa}{2} n_{\beta}) 
    - i\kappa r' \big|^{2}} \; = \; 
\frac{
\big|(q_{\alpha} -\frac{i\kappa}{2} n_{\alpha}) - 
     (q_{\beta} -\frac{i\kappa}{2} n_{\beta}) \big|^{2}}{
\big|(q_{\alpha} -\frac{i\kappa}{2} n_{\alpha}) - 
     (q_{\beta} -\frac{i\kappa}{2} n_{\beta})  - i\kappa n_{\beta} \big|^{2}} \; = \; 
\frac{\big|q_{\alpha}-q_{\beta} -\frac{i\kappa}{2}(n_{\alpha}-n_{\beta})\big|^{2}}{
      \big|q_{\alpha}-q_{\beta} -\frac{i\kappa}{2}(n_{\alpha}+n_{\beta})\big|^{2}}
\end{equation}
which proves Eq.(\ref{D1}).

\vspace{10mm}

Next, we substitute Eq.\ (\ref{D1}) into the expression Eq.\ (\ref{bas4-11})
for the partition function $Z(N,L)$ and represent it in the form Eq.\
(\ref{bas4-12}), where
\begin{eqnarray}
\nonumber
\tilde{Z}(N.L) &=&
 \int_{-\infty}^{+\infty} \frac{dq}{2\pi}\frac{1}{\kappa N} \;
\mbox{\LARGE e}^{-\frac{N L}{2\beta} q^{2} + \frac{\kappa^{2}L}{24\beta} N^{3} } \; +
\\
\nonumber
\\
\nonumber
&+& 
\sum_{M=2}^{\infty} \frac{1}{M!} \;  \sum_{n_{1}...n_{M}=1}^{\infty}
\biggl[\prod_{\alpha=1}^{M} 
\int_{-\infty}^{+\infty} \frac{d q_{\alpha}}{2\pi}\frac{1}{\kappa n_{\alpha}} \biggr]
\;{\boldsymbol \delta}\biggl(\sum_{\alpha=1}^{M} n_{\alpha}, \; N \biggr) \; 
\mbox{\LARGE e}^{-\frac{L}{2\beta}\sum_{\alpha=1}^{M} n_{\alpha} q_{\alpha}^{2} + 
\frac{\kappa^{2}L}{24\beta} \sum_{\alpha=1}^{M} n_{\alpha}^{3}} \; \times
\\
\nonumber
\\
&\times&
\prod_{\alpha<\beta}^{M} 
\frac{\big|q_{\alpha}-q_{\beta} -\frac{i\kappa}{2}(n_{\alpha}-n_{\beta})\big|^{2}}{
      \big|q_{\alpha}-q_{\beta} -\frac{i\kappa}{2}(n_{\alpha}+n_{\beta})\big|^{2}}
\label{D10}
\end{eqnarray}
Redefining all the momenta
\begin{equation}
\label{D11}
q_{\alpha} \; = \; \biggl(\frac{\beta\kappa}{L}\biggr)^{1/3}  p_{\alpha}
\end{equation}
and introducing a new parameter
\begin{equation}
\label{D12}
\lambda \; = \; \frac{1}{2} \biggl(\frac{L \kappa^{2}}{\beta}\biggr)^{1/3} \; = \; 
\frac{1}{2} \bigl(\beta^{5} u^{2} L\bigr)^{1/3}
\end{equation}
we rewrite:
\begin{equation}
\label{D13}
\frac{L}{2\beta} q_{\alpha}^{2} \; = \; \lambda p_{\alpha}^{2} \; ;
\end{equation}
\begin{equation}
\label{D14}
\frac{q_{\alpha}-q_{\beta} -\frac{i\kappa}{2}(n_{\alpha}-n_{\beta})}{
      q_{\alpha}-q_{\beta} -\frac{i\kappa}{2}(n_{\alpha}+n_{\beta})}
\; = \; 
\frac{p_{\alpha}-p_{\beta} -\lambda(n_{\alpha}-n_{\beta})}{
      p_{\alpha}-p_{\beta} -\lambda(n_{\alpha}+n_{\beta})}
\end{equation}
and obtain the integrals:
\begin{equation}
\label{D15}
\int_{-\infty}^{+\infty} \frac{d q_{\alpha}}{2\pi}\frac{1}{\kappa n_{\alpha}} 
\; \bigl[...\bigr] \; = \; 
\int_{-\infty}^{+\infty} \frac{d q_{\alpha}}{2\pi}\frac{\lambda}{\kappa}
\int_{0}^{+\infty} dt \;
\mbox{\LARGE e}^{-\lambda n_{\alpha} t} \; \bigl[...\bigr] \; = \; 
 \int_{-\infty}^{+\infty} \frac{d p_{\alpha}}{4\pi}
\int_{0}^{+\infty} dt \; 
\mbox{\LARGE e}^{-\lambda n_{\alpha} t} \; \bigl[...\bigr]
\end{equation}
Substituting the transformations, Eq.(\ref{D13})-(\ref{D15}), into 
Eq.(\ref{D10}), we get Eq.(\ref{bas4-13}).


\vspace{15mm}

\begin{center}
\appendix{\Large Appendix E} 
\end{center}

\newcounter{E}
\setcounter{equation}{0}
\renewcommand{\theequation}{E.\arabic{equation}}

\vspace{10mm}

To perform the summation over $n_{1},..., n_{M}$ in Eq.(\ref{bas4-17}), let us modify 
the pre-exponential factor:
\begin{equation}
\label{E1}
{\bf F} ({\bf p}; {\bf n}) \; \equiv \; 
\prod_{\alpha<\beta}^{M}
\frac{\big|p_{\alpha}-p_{\beta} -i\lambda(n_{\alpha}-n_{\beta})\big|^{2}}{
      \big|p_{\alpha}-p_{\beta} -i\lambda(n_{\alpha}+n_{\beta})\big|^{2}} 
\; = \; 
\prod_{\alpha\not=\beta}^{M}
\frac{\bigl(|p_{\alpha}-p_{\beta}| + i\lambda(n_{\alpha}-n_{\beta})\bigr)}{
      \sqrt{\bigl(|p_{\alpha}-p_{\beta}| + i\lambda(n_{\alpha}+n_{\beta})\bigr)
            \bigl(|p_{\alpha}-p_{\beta}| - i\lambda(n_{\alpha}+n_{\beta})\bigr)}}
\end{equation}
We introduce the auxiliary fields $\chi_{\alpha\beta}$, $\phi_{\alpha\beta}$,
and $\psi_{\alpha\beta}$ to generate the numerator 
\begin{equation}
\label{E2}
 \bigl(|p_{\alpha}-p_{\beta}| + i\lambda(n_{\alpha}-n_{\beta})\bigr) 
\; = \;
\frac{\partial}{\partial\chi_{\alpha\beta}}
\exp\bigl[
\bigl(|p_{\alpha}-p_{\beta}| + i\lambda(n_{\alpha}-n_{\beta})\bigr) \chi_{\alpha\beta}
\bigr]\bigg|_{\chi_{\alpha\beta}=0}
\end{equation}
and the denominators
\begin{eqnarray}
\label{E3}
\frac{1}{\sqrt{\bigl(|p_{\alpha}-p_{\beta}| + i\lambda(n_{\alpha}+n_{\beta})\bigr)}}
&=& 
\int_{-\infty}^{+\infty} \frac{d\phi_{\alpha\beta}}{\sqrt{2\pi}}
\exp\biggl[-\frac{1}{2} \bigl( |p_{\alpha}-p_{\beta}|  
+ i \lambda (n_{\alpha}+n_{\beta}) \bigr) \phi_{\alpha\beta}^{2} \biggr]
\\
\nonumber
\\
\label{E4}
\frac{1}{\sqrt{\bigl(|p_{\alpha}-p_{\beta}| - i\lambda(n_{\alpha}+n_{\beta})\bigr)}}
&=& 
\int_{-\infty}^{+\infty} \frac{d\psi_{\alpha\beta}}{\sqrt{2\pi}}
\exp\biggl[-\frac{1}{2} \bigl( |p_{\alpha}-p_{\beta}| 
- i \lambda (n_{\alpha}+n_{\beta})\bigr) \psi_{\alpha\beta}^{2} \biggr]
\end{eqnarray}
Combining together Eqs.(\ref{E2})-(\ref{E4}) we rewrite the prefactor as
\begin{equation}
\label{E5}
{\bf F} ({\bf p}; {\bf n}) 
\; = \; 
\biggl( \prod_{\alpha\not=\beta}^{M} 
\iint_{-\infty}^{+\infty} \frac{d\psi_{\alpha\beta}d\phi_{\alpha\beta}}{2\pi} 
\frac{\partial}{\partial\chi_{\alpha\beta}} \biggr)
\exp\biggl[
-\frac{1}{2} \sum_{\alpha\not=\beta}^{M}
|p_{\alpha}-p_{\beta}| 
(\phi_{\alpha\beta}^{2} + \psi_{\alpha\beta}^{2} - 2\chi_{\alpha\beta})
+ i \lambda \sum_{\alpha=1}^{M} \eta_{\alpha} n_{\alpha} 
\biggr]\Bigg|_{\{\chi_{\alpha\beta}\}=0}
\end{equation}
where
\begin{equation}
   \label{E6}
\eta_{\alpha} \; \equiv \; 
\eta_{\alpha}({\boldsymbol \psi}, {\boldsymbol \phi}, {\boldsymbol \chi}) 
\; = \; \frac{1}{2} \sum_{\beta\not=\alpha}^{M}
\biggl(\psi_{\alpha\beta}^{2} + \psi_{\beta\alpha}^{2} - \phi_{\alpha\beta}^{2} - \phi_{\beta\alpha}^{2}
+2 \chi_{\alpha\beta} -2 \chi_{\beta\alpha} \biggr)
\end{equation}
Introducing now the integro-differential operator
\begin{equation}
\label{E7}
\int \hat {\cal G}_{M}({\bf p}; \; {\boldsymbol \psi}, {\boldsymbol \phi}, {\boldsymbol \chi}) 
\; \equiv \; 
\biggl[\prod_{\alpha\not=\beta}^{M} 
\iint_{-\infty}^{+\infty} \frac{d\psi_{\alpha\beta}d\phi_{\alpha\beta}}{2\pi} 
\frac{\partial}{\partial\chi_{\alpha\beta}} \biggr]\; 
\mbox{\LARGE e}^{-\frac{1}{2}\sum_{\alpha\not=\beta}^{M}\big|p_{\alpha}-p_{\beta}\big|
(\psi_{\alpha\beta}^{2}+\phi_{\alpha\beta}^{2} - 2\chi_{\alpha\beta}) }
\end{equation}
as well as the integration operator
\begin{equation}
   \label{E8}
\int {\cal D}_{M}({\bf y,p}) \; \equiv \;
\prod_{\alpha=1}^{M} \int_{-\infty}^{+\infty} dy_{\alpha}
\int_{-\infty}^{+\infty} \frac{dp_{\alpha}}{4\pi} \int_{0}^{+\infty} dt_{\alpha} 
\Ai(y_{\alpha}+t_{\alpha}+p^{2}_{\alpha}) 
\end{equation}
the partition function, Eq.(\ref{bas4-17}), can be represented as follows
\begin{eqnarray}
\nonumber
\tilde{Z}(N,\lambda) 
&=& 
 \int {\cal D}_{1}(y,p) \; \exp(\lambda N y) \; + \\
&+& \sum_{M=2}^{\infty} \frac{1}{M!} 
\int {\cal D}_{M}({\bf y,p})
\int \hat {\cal G}_{M}({\bf p}; \; {\boldsymbol \psi}, {\boldsymbol \phi}, {\boldsymbol \chi}) 
\sum_{n_{1}...n_{M}=1}^{\infty} 
{\boldsymbol \delta}\biggl(\sum_{\alpha=1}^{M}n_{\alpha}, \; N\biggr)
\prod_{\alpha=1}^{M}
\mbox{\LARGE e}^{\lambda (y_{\alpha} + i \eta_{\alpha}) n_{\alpha}}
\label{E9}
\end{eqnarray}

\vspace{5mm}

Next we prove that 
\begin{eqnarray}
\nonumber
S(N,M) &\equiv& \sum_{n_{1}...n_{M}=1}^{\infty} 
 a_{1}^{n_{1}} \, a_{2}^{n_{2}} \, a_{3}^{n_{3}} \, ... \, a_{M}^{n_{M}} \; 
{\boldsymbol \delta}\biggl(\sum_{\alpha=1}^{M}n_{\alpha}, \; N\biggr)  \; = 
\\
\nonumber 
\\
\nonumber
&=&
a_{1}^{N}  
\frac{a_{2}}{(a_{1}-a_{2})}
\frac{a_{3}}{(a_{1}-a_{3})}  ... 
\frac{a_{M}}{(a_{1}-a_{M})}
\; + \; 
a_{2}^{N} 
\frac{a_{1}}{(a_{2}-a_{1})}  
\frac{a_{3}}{(a_{2}-a_{3})}  ...  
\frac{a_{M}}{(a_{2}-a_{M})}
\; + \; 
\\
\nonumber \\
&+&
a_{3}^{N} 
\frac{a_{1}}{(a_{3}-a_{1})}  
\frac{a_{2}}{(a_{3}-a_{2})}  ... 
\frac{a_{M}}{(a_{3}-a_{M})}
\; + \; ... \; + \; 
a_{M}^{N}  
\frac{a_{1}}{(a_{M}-a_{1})} 
\frac{a_{2}}{(a_{M}-a_{2})}  ...  
\frac{a_{M-1}}{(a_{M}-a_{M-1})}
\label{E10}
\end{eqnarray}
To unbound the summations over $n_{1}, n_{2}, ..., n_{M}$ in the above series, 
let us introduce the integral representation of the Kronecker symbol:
\begin{equation}
 \label{E11}
{\boldsymbol \delta}(k, \; m)  \; = \; 
\oint_{\cal C} \; \frac{dz}{2\pi i} \; z^{m - k - 1}
\end{equation}
where both $k$ and $m$ are assumed to be positive integers and 
the integration in the complex plane goes over a closed contour ${\cal C}$ 
around zero. Using this representation we obtain
\begin{equation}
\label{E12}
S(N,M) \; = \; 
\oint_{\cal C} \; \frac{dz}{2\pi i} \; z^{N-1} \; 
\prod_{\alpha=1}^{M} \sum_{n_{\alpha}=1}^{\infty} 
\biggl(\frac{a_{\alpha}}{z}\biggr)^{n_{\alpha}}
\end{equation}
where it is assumed that the radius $R_{\cal C}$ of the the contour ${\cal C}$ is big enough,
namely $R_{\cal C} > \max_{\alpha} |a_{\alpha}|$, so that all the complex numbers 
$a_{1}, a_{2}, ..., a_{M}$ are contained inside ${\cal C}$. 
In this case all the summations in Eq.(\ref{E12}) are convergent, and we find
\begin{equation}
\label{E13}
S(N,M) \; = \; 
\oint_{\cal C} \; \frac{dz}{2\pi i} \; z^{N-1} \; 
\frac{a_{1}}{(z - a_{1})} \cdot \frac{a_{2}}{(z - a_{2})} \cdot \; ... \; \cdot
\frac{a_{M}}{(z - a_{M})}
\end{equation}
Since the above integral is equal to the sum of $M$ pole contributions at 
$z = a_{\alpha} \; (\alpha = 1, 2, ... M)$, one gets Eq.(\ref{E10}).

Now substituting Eq.(\ref{E10}) 
(with $a_{\alpha} = \exp\bigl[\lambda(y_{\alpha} + i\eta_{\alpha})\bigr]$) 
into Eq.(\ref{E9}) we obtain
\begin{eqnarray}
\nonumber
\tilde{Z}(N,\lambda) 
&=& 
 \int {\cal D}_{1}(y,p) \; \exp(\lambda N y) \; + \\
&+& \sum_{M=2}^{\infty} \frac{1}{M!} \; M 
\int {\cal D}_{M}({\bf y,p})
\int \hat {\cal G}_{M}({\bf p}; \; {\boldsymbol \psi}, {\boldsymbol \phi}, {\boldsymbol \chi}) 
\; \mbox{\LARGE e}^{\lambda N (y_{1} + i \eta_{1})}
\prod_{\alpha=2}^{M}
\frac{
\mbox{\LARGE e}^{\lambda (y_{\alpha} + i \eta_{\alpha})}}{
\mbox{\LARGE e}^{\lambda (y_{1} + i \eta_{1})} \; - \; 
\mbox{\LARGE e}^{\lambda (y_{\alpha} + i \eta_{\alpha})}}
\label{E14}
\end{eqnarray}
where we have used the symmetry of the expressions in Eqs.(\ref{E7}), (\ref{E8}) with respect
to permutations of $\{(y_{\alpha} + i \eta_{\alpha})\}$.


\vspace{15mm}

\begin{center}
\appendix{\Large Appendix F} 
\end{center}

\newcounter{F}
\setcounter{equation}{0}
\renewcommand{\theequation}{F.\arabic{equation}}

\vspace{10mm}

In this Appendix, we show that, independently of the values of the
momenta $p_{\alpha}$ and the parameter $s$, the integro-differential
factor in Eq.\ (\ref{bas5-3}) provides unity,
\begin{equation}
   \label{F1}
{\boldsymbol {\cal G}} \; \equiv \; 
\int \hat {\cal G}_{M}({\bf p}; \; {\boldsymbol \psi}, {\boldsymbol \phi}, {\boldsymbol \chi}) \; 
\mbox{\LARGE e}^{i s  \eta_{1}({\boldsymbol \psi}, {\boldsymbol \phi}, {\boldsymbol \chi})} 
\; = \; 1
\end{equation}
Substituting here the definitions, Eqs(\ref{E6}) and (\ref{E7}), the factor ${\boldsymbol {\cal G}}$
can be split into two factors
\begin{equation}
 \label{F2}
{\boldsymbol {\cal G}} \; = \; {\boldsymbol {\cal G}}_{A} \; {\boldsymbol {\cal G}}_{B}
\end{equation}
where
\begin{equation}
 \label{F3}
{\boldsymbol {\cal G}}_{A} \; = \;
 \biggl[\prod_{2\leq\alpha\not=\beta}^{M} 
\iint_{-\infty}^{+\infty} \frac{d\psi_{\alpha\beta}d\phi_{\alpha\beta}}{2\pi} 
\frac{\partial}{\partial\chi_{\alpha\beta}} \biggr]\; 
\mbox{\LARGE e}^{-\frac{1}{2}\sum_{\alpha\not=\beta}^{M}\big|p_{\alpha}-p_{\beta}\big|
(\psi_{\alpha\beta}^{2}+\phi_{\alpha\beta}^{2} - 2\chi_{\alpha\beta})}\Bigg|_{\{\chi_{\alpha\beta}\}=0}
\end{equation}
and
\begin{eqnarray}
 \nonumber
{\boldsymbol {\cal G}}_{B} 
&=&
 \biggl[\prod_{\alpha=2}^{M} 
\iint_{-\infty}^{+\infty} \frac{d\psi_{\alpha 1}d\phi_{\alpha 1}}{2\pi} 
\frac{\partial}{\partial\chi_{\alpha 1}} 
\iint_{-\infty}^{+\infty} \frac{d\psi_{1\alpha}d\phi_{1\alpha}}{2\pi} 
\frac{\partial}{\partial\chi_{1\alpha}}\biggr] \times 
\\
\nonumber
\\
\nonumber
&\times&
\exp\biggl[
-\frac{1}{2}\sum_{\alpha=2}^{M}\big|p_{\alpha}-p_{1}\big| 
(\psi_{\alpha 1}^{2} + \phi_{\alpha 1}^{2} + \psi_{1\alpha}^{2} + \phi_{1\alpha}^{2}) + 
\sum_{\alpha=2}^{M}\big|p_{\alpha}-p_{1}\big| (\chi_{\alpha 1} + \chi_{1 \alpha} ) \; +
\\
&& \; + \; 
\frac{i}{2} s \sum_{\alpha=2}^{M} 
(\psi_{\alpha 1}^{2} - \phi_{\alpha 1}^{2} + \psi_{1\alpha}^{2} - \phi_{1\alpha}^{2}) 
\; + \; 
i s \sum_{\alpha=2}^{M} (\chi_{1\alpha} - \chi_{\alpha 1} ) 
\biggr]\Bigg|_{\{\chi_{\alpha 1},\chi_{1\alpha}\}=0}
\label{F4}
\end{eqnarray}
For the factor ${\boldsymbol {\cal G}}_{A}$, Eq.(\ref{F3}), we easily find
\begin{equation}
 \label{F5}
{\boldsymbol {\cal G}}_{A} \; = \;
 \prod_{2\leq\alpha\not=\beta}^{M} 
\frac{1}{\sqrt{\big|p_{\alpha}-p_{\beta}\big|}} \cdot 
\frac{1}{\sqrt{\big|p_{\alpha}-p_{\beta}\big|}} \cdot 
\big|p_{\alpha}-p_{\beta}\big| \; \equiv \; 1
\end{equation}
In a similar way we obtain for the factor ${\boldsymbol {\cal G}}_{B}$
\begin{eqnarray}
 \nonumber
{\boldsymbol {\cal G}}_{B} 
&=&
\prod_{\alpha=2}^{M}
\int_{-\infty}^{+\infty} \frac{d\psi_{\alpha 1}}{\sqrt{2\pi}} 
\exp\biggl(-\frac{1}{2}\bigl[|p_{\alpha}-p_{1}| - is\bigr] \psi_{\alpha 1}^{2}\biggr)
\times  
\prod_{\alpha=2}^{M}
\int_{-\infty}^{+\infty} \frac{d\phi_{\alpha 1}}{\sqrt{2\pi}} 
\exp\biggl(-\frac{1}{2}\bigl[|p_{\alpha}-p_{1}| + is\bigr] \phi_{\alpha 1}^{2}\biggr) 
\times
\\
\nonumber
\\
\nonumber
&\times&
\prod_{\alpha=2}^{M}
\int_{-\infty}^{+\infty} \frac{d\psi_{1\alpha}}{\sqrt{2\pi}} 
\exp\biggl(-\frac{1}{2}\bigl[|p_{\alpha}-p_{1}| - is\bigr] \psi_{1\alpha}^{2}\biggr)
\times  
\prod_{\alpha=2}^{M}
\int_{-\infty}^{+\infty} \frac{d\phi_{1\alpha}}{\sqrt{2\pi}} 
\exp\biggl(-\frac{1}{2}\bigl[|p_{\alpha}-p_{1}| + is\bigr] \phi_{1\alpha}^{2}\biggr) 
\times
\\
\nonumber
\\
\label{F6}
&\times&
\prod_{\alpha=2}^{M}
\frac{\partial}{\partial\chi_{\alpha 1}} 
\exp\biggl(\bigl[|p_{\alpha}-p_{1}| - i s \bigr]\chi_{\alpha 1}\biggr)\Bigg|_{\chi_{\alpha 1}=0} 
\times
\prod_{\alpha=2}^{M}
\frac{\partial}{\partial\chi_{1\alpha}} 
\exp\biggl(\bigl[|p_{\alpha}-p_{1}| + i s \bigr]\chi_{1\alpha}\biggr)\Bigg|_{\chi_{1\alpha}=0} 
\; \;  
\\
\nonumber
\\
\nonumber
\\
\label{F7}
&=&
\prod_{\alpha=2}^{M}
\frac{1}{(p_{\alpha}-p_{1})^{2} + s^{2}} \times 
\bigl[(p_{\alpha}-p_{1})^{2} + s^{2}\bigr] \; \equiv \; 1
\end{eqnarray}
which proves Eq.(\ref{F1}).



\end{document}